\documentstyle[prb,eqsecnum,aps,epsf]{revtex}
\begin{document}
\draft
\title{
Theory of Tunneling Anomaly in Superconductor above Paramagnetic Limit
}
\author{
Hae-Young Kee$^{1}$, I.L. Aleiner$^{2,3}$, and B.L. Altshuler$^{2,4}$}
\address{
$^{1}$Serin Physics Laboratory, Rutgers University, Piscataway, NJ 08855\\
$^{2}$NEC Research Institute,
4 Independence Way, Princeton, NJ 08540\\
$^{3}$Department of Physics and Astronomy,
SUNY at Stony Brook, Stony Brook, NY 11794
\\
$^{4}$Physics Department, Princeton University, Princeton, NJ 08544
}
\maketitle

\begin{abstract}
We study the tunneling density of states (DoS) in the superconducting
systems driven by Zeeman splitting $E_Z$ into the paramagnetic
phase. We show that, even though the BCS gap disappears,
superconducting fluctuations cause a strong DoS singularity in the
vicinity of energies $-E^\ast$ for electrons polarized along the
magnetic field and $E^\ast$ for the opposite polarization.  The
position of this singularity $E^\ast=\case{1}{2}\left(E_Z +
\sqrt{E_Z^2- \Delta^2} \right)$ (where $\Delta$ is BCS gap at $E_Z=0$)
is universal. We found analytically the shape of the DoS for different
dimensionality of the system. For ultra-small grains the singularity
has the shape of the hard gap, while in higher dimensions it appears
as a significant though finite dip.  The spin-orbit scattering, and
the orbital magnetic field suppress the singularity.  Our results are
qualitatively consistent with recent experiments in superconducting
films.  
\end{abstract}
\pacs{PACS numbers: 74.40.+k, 74.50.+r, 73.40Gk, 73.50.-h}

\section{Introduction}
It is well known that the magnetic field, $H$, suppresses
superconductivity since it lifts time reversal symmetry (see {\em
e.g.} Ref.~\onlinecite{Tinkham} for general introduction).  In the
absence of the spin orbit coupling, this effect can be separated into
two mechanisms: (i) effect of the magnetic field on the orbital motion
associated with Aharonov-Bohm phase, and (ii) Zeeman splitting of the
states with the same spatial wave functions but opposite spin
directions.

In the bulk systems, the suppression of the superconductivity is
typically associated with the first mechanism.  Indeed, the estimate
for the critical field $H_{c_2}$ in this case is
\begin{equation}
H_{c_2}\xi^2 \simeq \phi_0,
\label{Horbit}
\end{equation}
where $\phi_0=\frac{hc}{2e}$ is the superconducting flux quantum and 
\begin{equation}
\xi = \sqrt{\frac{D}{\Delta}}
\label{scgap}
\end{equation}
is the coherence length for the dirty superconductors,
$\Delta$ is the BCS gap, and $D$ is the diffusion
coefficient. On the other hand, the magnetic field necessary to affect
the superconductivity by virtue of the spin mechanism is given by
\begin{equation}
g_L\mu_B H_{spin} \simeq \Delta,
\label{Hspin}
\end{equation}
where $g_L$ is the Land\'{e} $g$-factor, and $\mu_B=\frac{e\hbar}{2mc}$ is the 
Bohr magneton.
Comparing Eqs. (\ref{Horbit}) and (\ref{Hspin}), one finds that
$H_{spin}$ is far in excess of $H_{c_2}$:
\begin{equation}
\frac{H_{spin}}{H_{c_2}} \simeq \epsilon_F \tau \gg 1,
\label{ratio}
\end{equation}
where $\epsilon_F$ is the Fermi energy and $\tau$ is the elastic
momentum relaxation time.
Condition (\ref{ratio}) means that in the bulk system, the orbital
effect of the magnetic field is always dominant.

Situation may change in the restricted geometries.  Consider, {\em
e.g.}, the superconducting film of the thickness $a \ll \xi$, placed
in the magnetic field parallel to the plane of the film.  The Cooper
pair in this case is restricted in the transverse direction by the
film thickness $a$.  As a result, the geometrical area swept by this
pair can be estimated as $a\xi$ rather than as $\xi^2$.  Therefore,
Eq. (\ref{Horbit}) should be changed to
\begin{equation}
H_{c_2}^{\parallel} \xi a \simeq \phi_0\quad \Rightarrow \quad 
{H_{c_2}^{\parallel}} \simeq H_{c_2}\left( \frac{\xi}{a} \right).
\end{equation}
On the other hand, Zeeman splitting $E_Z=g_L \mu_B H$ is not affected by
geometrical restriction.
Accordingly, instead of Eq. (\ref{ratio}) the ratio of the two scales of
magnetic field is given by
\begin{equation}
\frac{H_{spin}}{H_{c_2}^{\parallel}} \simeq (\epsilon_F \tau) 
\left( \frac{a}{\xi} \right).
\label{ratio2}
\end{equation}
Thus, for sufficiently thin films, $a \ll \frac{\xi}{\epsilon_F \tau}$,
the spin effects become dominant.
One can easily check that the same estimate (\ref{ratio2}) holds
for other restricted systems, i.e., superconducting grains or wires.
In these cases, $a$ is the size of the grain or the diameter of the wire
respectively.
Quite generally, $a$ is determined by the minimal size of the sample
in the plane perpendicular to the magnetic field.
In this paper we consider such restricted geometries, and unless the
opposite is stated, neglect the orbital effects.

The transition from superconductor to paramagnet is of the first
order\cite{ClCh}: superconducting state is the only stable state at
$E_Z \leq \Delta$; while at $E_Z \geq 2\Delta$ normal state is the
only stable state.
Both phases are locally stable in the interval of magnetic fields
where $\Delta < E_Z < 2\Delta$.  The normal state becomes
lowest in energy and thus globally
stable at $E_Z \geq \sqrt{2}\Delta$. From now on, we will assume that
this condition is fulfilled.

One of the most fundamental manifestations of the superconductivity is
the gap in the tunneling density of states (DoS) around the zero
bias\cite{Tinkham,Giaver}.  
One can expect that after the paramagnetic transition not only BCS
order parameter vanishes but also
the energy dependence of tunneling DoS
becomes similar to those in superconductors above critical
temperature $T_c$. (The latter dependence is discussed in the review
article Ref. \onlinecite{AA}.)

In this paper, we demonstrate that, on the contrary, there are clear
observable  superconducting effects in the normal state even far from
the transition region.
We will show that at the transition point there appears a dip in the
DoS (schematic evaluation of the DoS with the magnetic field is shown in
Fig.~\ref{Fig:1}).

\narrowtext
{\begin{figure}[ht]
\vspace{0.2 cm}
\epsfxsize=7.7cm
\centerline{\epsfbox{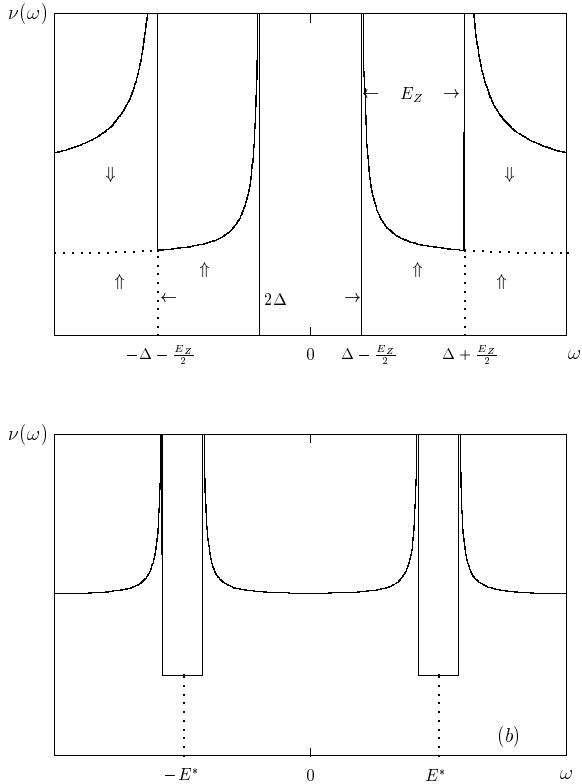}}
\vspace{0.2cm}
\caption{Evolution of the tunneling DoS with the Zeeman splitting
$E_Z$ for (a) the superconducting state, $E_Z < \protect\sqrt{2}\Delta
$, 
see {\em e.g.} Ref.~\protect\onlinecite{Fulde},
and
(b) for the paramagnetic state $E_Z > \protect\sqrt{2}\Delta$.  
The usual zero
bias anomaly in the paramagnetic state (b) is not shown for
simplicity. Shape of the singularity at $E^*$ corresponds to the
$0$-dimensional case.
}
\label{Fig:1}
\end{figure} 
}
\widetext

The shape and the width of this dip depend on dimensionality of the
system. However, its position is remarkably universal:
\begin{equation}
E^*=\frac{1}{2}(E_Z+\sqrt{E_Z^2-\Delta^2}),
\label{E^*}
\end{equation}
for 0D (grain), 1D (strip), and 2D (film) cases.  Some of conclusions
have been already briefly reported by two of us\cite{Igor}.  Here we
present detailed derivations of results of Ref.~\onlinecite{Igor} and
consider the relevant perturbations (spin-orbital coupling, orbital
magnetic field, finite temperature, and energy relaxation) of the new
tunneling anomaly.

The remainder of the paper is organized as
follows. Section~\ref{sec:0d} presents the parametrically exact
solution for the simplest but instructive geometry of
zero-dimensional systems (ultra-small superconducting
grains). Section~\ref{sec:hd} deals with more involved problem of the
tunneling anomaly in the  superconducting films and wires. Both
sections required application of the diagrammatic technique on the level
of at least Ref.~\onlinecite{AGD}. For the benefit of the readers
interested in physical interpretation rather than in  
rigorous derivations, we present in Sec.~\ref{sec:qd} the
qualitative derivation which grasps all the essential physics
involved, even though fails to give the completely quantitative
description. Section~\ref{sec:rp} analyzes how the tunneling anomaly
is affected by spin-orbital coupling, orbital magnetic field, finite
temperature, and energy relaxation. We discuss the recent
experiment\cite{WWA} on the Zeeman splitting of the tunneling anomaly
in $Al$ films in Sec.~\ref{sec:exp}.  Our findings are summarized in
Conclusion.

\section{ Zero  dimensional systems}
\label{sec:0d}

Let us consider an isolated disordered superconducting grain which is
small so that  the Zeeman splitting dominates over the orbital magnetic
field effect (see, {\em e.g}, Refs.~\onlinecite{Ralph,Ralph1} for recent
experiments on such grains). 
We assume that the size of the grain exceeds electronic mean free path
$l$, and, at the same time, it is much smaller than the superconducting
coherence length $\xi$. We also assume that $k_Fl \gg 1$. This results
in the large dimensionless conductance of the grain
$g\ (g \sim k_F^2 l a)$. Finally, we assume that the grain is already
driven into the paramagnetic state by  the Zeeman splitting. Our goal
is to find  effects of the superconducting fluctuations on the DoS
of the system.

The  Hamiltonian $H$ of the system consists of noninteracting part $H_0$ 
and interacting one $H_{int}$. 
Using the basis of the exact eigenstates of $H_0$ 
labeled by integers $i$ and $j$, one can write the Hamiltonian as
\begin{equation}
H = \sum_{i \sigma} E_{i \sigma} a^{\dagger}_{i \sigma} a_{i \sigma}
-\lambda \bar{\delta}
\sum_{i, j} a^{\dagger}_{i \uparrow} a^{\dagger}_{i \downarrow}
a_{j \downarrow} a_{j \uparrow}.\label{ham}
\end{equation}
Here operator $a^{\dagger}_{i \sigma} (a_{i \sigma})$
creates(annihilates) an electron in a state $i$ with spin $\sigma=
\uparrow, \downarrow$, and energy $E_{i \uparrow (\downarrow)} =
\epsilon_i \mp E_Z/2$ where $\epsilon_i$ is the orbital energy of
$i$-th state.  $\lambda \ll 1$ is the dimensionless interaction
constant, and $\bar{\delta}$ is the average level spacing:
\begin{equation}
\langle \epsilon_{i+1}-\epsilon_i \rangle = \bar{\delta}.
\end{equation}

Let us stop for a moment to discuss the approximation made in Eq.~(\ref{ham}). 
We included in Eq. (\ref{ham}) only the matrix elements of the interaction
Hamiltonian responsible for the superconductivity.
We omitted two kinds of diagonal terms.  
The term proportional to 
$ a^{\dagger}_{i \sigma_1} a_{i \sigma_1} a^{\dagger}_{j \sigma_2}
a_{j \sigma_2}$ represents 
the total charging energy responsible for the Coulomb blockade\cite{CB}.
It is not important for us because it does not lead to any anomalies
at energies of the order of Zeeman splitting, and it can be 
accounted for by the corresponding shift of the applied bias.
Other diagonal terms such as the one  proportional to 
$ a^{\dagger}_{i \sigma_1} a_{i \sigma_2} a^{\dagger}_{j \sigma_2}
a_{j \sigma_1}$ represents the spin exchange.
It is not included because it leads only to the
renormalization of the Land\'{e} factor $g_L$.

We also omitted off-diagonal terms, such as
$ a^{\dagger}_{i \uparrow} a^{\dagger}_{j \downarrow}
a_{k \downarrow} a_{l \uparrow}$ with $i$, $j$, $k$, $l$ not equal
pairwise, corresponding to the matrix elements:
\[
M^{kl}_{ij} = \int dr dr^{\prime}  V(r-r^{\prime}) \psi^*_i(r) 
\psi^*_j(r^{\prime}) \psi_k(r) \psi_l(r^{\prime}).
\]
The wave functions are known to oscillate
very fast, so the wavefunctions of different levels
are very weakly correlated.
We can restrict our consideration by
short range interaction, $V(r-r^{\prime}) = \frac{\lambda}{\nu_0}
\delta(r-r^{\prime})$, where $\nu_0$ is the bare DoS.
One see that the integrand ($|\psi_i(r)|^2 |\psi_j(r)|^2$)
in the diagonal matrix elements is always positive  
while the product ($\psi^*_i(r) \psi^*_j(r) \psi_k(r) \psi_l(r)$) 
can be both positive and negative.
As a result, the off-diagonal matrix elements turn out to be smaller
than diagonal ones. 
Straightforward calculation\cite{1overg,Zinoviev} 
shows that they are smaller by the factor
$1/g$.

In the paramagnetic state ($E_Z > \sqrt{2} \Delta$),
the structure of the ground state is similar to that without
interaction, see Fig.~\ref{Fig:2.1}.
The orbitals with $\epsilon_i < -E_Z/2$ are doubly
occupied while those with $\epsilon_i > E_Z/2$ are empty.
The orbitals with $|\epsilon_i| < E_Z/2$ are
spin-polarized with up-spin.

\narrowtext
{\begin{figure}[ht]
\vspace{0.2cm}
\epsfxsize=6cm
\hspace*{0.5cm}
\epsfbox{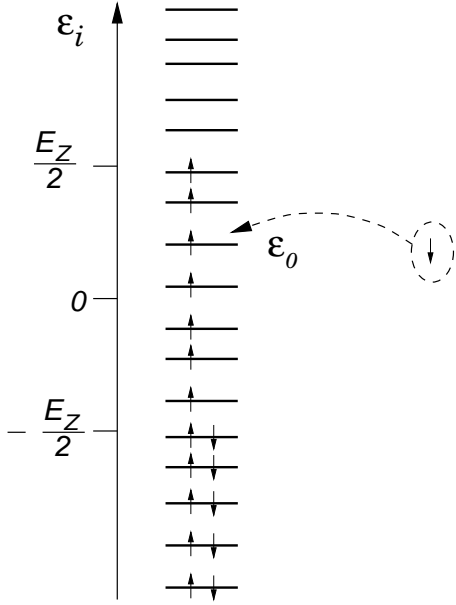}
\vspace{0.9 cm}
\caption{
Structure of the ground state of the superconductor above paramagnetic
limit. Electron tunneling onto the orbital $\epsilon_0$, creates the
spin single states on this orbital. At some energy $\epsilon_0$ mixing
of this singlet with the empty states becomes resonant, see the text.
}
\label{Fig:2.1}
\end{figure} 
}
\widetext

The Hamiltonian (\ref{ham}) does not affect the spin polarized states, but
mixes the doubly-occupied and empty states.
Since those states are separated from each other by a large gap $E_Z$,
this mixing can be treated perturbatively.
Thus, the mixing does not change the ground state qualitatively.
On the contrary, the spectrum of the excitations, i.e., the tunneling 
DoS changes drastically due to the interaction.
The essence of this effect is that spin-down electron tunneling into some
orbital $\epsilon_0$ already occupied by spin down electron creates
an electron pair which can mix with the empty orbitals and thus
interact with superconducting fluctuations.
This mixing turns out to be resonant at some energy $E=E^*$ and
it leads to the sharp singularity in the spectrum of one-electron excitation.

To evaluate the effect of superconducting fluctuations on the DoS
of electrons in the paramagnetic state, we use
the diagrammatic technique for the Green function (GF) 
at zero-temperature\cite{AGD}.
The DoS can be expressed through the one particle GF,
$G_{i\sigma}(\omega)$,
of an electron on the orbital $j$ and with spin $\sigma=\pm 1\equiv\uparrow
(\downarrow)$:
\begin{equation}
\nu_{\sigma}(\omega)= - \frac{1}{\pi} sgn(\omega) 
{\rm Im} \sum_i G_{i\sigma}(\omega),\label{dos}
\end{equation}
where
\begin{equation}
G^{-1}_{i \sigma}= G^{0 -1}_{i \sigma}-\Sigma_{i\sigma}.\label{dyson}
\end{equation}
$G^0_{i\sigma}$ is the GF for the non-interacting system
\begin{equation}
G^0_{i\uparrow(\downarrow)}=(\omega_+ -\epsilon_i \pm E_Z/2)^{-1},
\label{G_0}
\end{equation}
and $\Sigma_{i\sigma}$ is the one particle self-energy.

Leading contribution to the self-energy is shown in
Fig. \ref{Fig:2.2}a.  The solid and curly lines denote 
single-particle GF and the propagator of superconducting fluctuations,
respectively.  The latter can be obtained by summing the polarization
loops in the Cooper channel shown in Fig. \ref{Fig:2.2}b. 
The single loop is given by
\begin{equation}
\Pi(\omega)= \frac{1}{2\bar{\delta}} {\rm ln} \left( \frac{\omega_c^2}
{E_Z^2-\omega^2_+ } \right),
\end{equation}
where $\omega_+ = \omega + i 0 sgn(\omega)$ and $\omega_c$ is the
high-energy cut-off.  Solving the Dyson equation
(Fig.~\ref{Fig:2.2}b), we obtain the propagator
\begin{equation}
\Lambda(\omega)= \frac{\lambda \bar{\delta}}
{1-\lambda \bar{\delta} \Pi(\omega) } 
=\frac{2 \bar{\delta} }{ {\rm ln} \left(
\frac{E_Z^2-\omega_+^2}{\Delta^2} \right) }, 
\label{Lam0}
\end{equation}
where $\Delta=\omega_c \exp{(-1/\lambda)} $  is the BCS gap.

The propagator (\ref{Lam0}) has the pole at $\omega= \pm \Omega$,
\begin{equation}
\Omega = \sqrt{E_Z^2-\Delta^2}. \label{omega}
\end{equation}
This pole can be interpreted as the bound state of two
quasiparticles with energy $\Omega$.

\narrowtext
{\begin{figure}[ht]
\vspace{0.1cm}
\epsfxsize=7cm
\hspace*{0.5cm}
\epsfbox{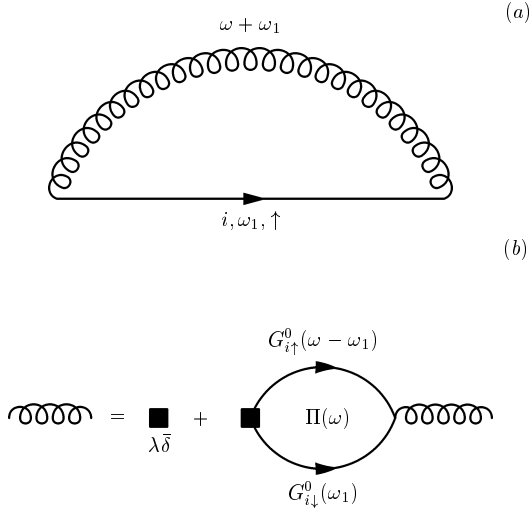}
\vspace{-0.05 cm}
\caption{
Diagrams for (a) self-energy $\Sigma_{i,\downarrow}(\omega )$ and (b)
superconducting propagator $\Lambda(\omega)$.
}
\label{Fig:2.2}
\end{figure} 
}
\widetext

Analytic expression for the self-energy given by diagram Fig.~\ref{Fig:2.2}a,
has the form
\begin{equation}
\Sigma_{i \downarrow}(\omega) 
= i \int^{\infty}_{-\infty} \frac{d\omega_1}{2\pi}
\Lambda(\omega+\omega_1) G^0_{i\uparrow}(\omega_1).
\end{equation}
One can see that there are two contributions
to the self-energy.
One comes from the pole of $\Lambda$ and the other is due to the branch-cut
of this propagator.
 The pole contribution gives a singularity of 
the self-energy at certain $\omega$ and $\epsilon_i$
while the contribution of the branch-cut is smooth.
To find  the singularity in the DoS, only  the pole
contribution to $\Sigma$ may be retained:
\begin{equation}
\Sigma_{i \downarrow}(\omega) 
= \frac{\bar{\delta} \Delta^2}{\Omega} \frac{1}{\omega_++\epsilon_i-E_Z/2
+\Omega sgn(\epsilon_i-E_Z/2)}.\label{self}
\end{equation}

At certain $\omega$ the pole of the self-energy coincides with the pole of
$G^0$. This causes the singularity in the DoS.
One can check that singularities of Eq. (\ref{G_0}) and Eq. (\ref{self})
coincide provided $\epsilon_i =\Omega/2$ and
\begin{equation}
 \omega = \frac{E_Z+\Omega}{2}=\frac{E_Z+\sqrt{E_Z^2-\Delta^2}}{2}
\equiv E^*. \label{singular}
\end{equation}

Substituting  Eq. (\ref{self}) into Eq. (\ref{dyson})
we obtain the GF for the down-spin electron at $\omega$ close to
$E^*$:
\begin{equation}
G_{i\downarrow}(\omega) = \frac{\omega_+ + \epsilon_i-E_Z/2-\Omega}
{(\omega_+ -\epsilon_i-E_Z/2)(\omega_+ +\epsilon_i-E_Z/2-\Omega)
-W_0^2 },\label{GF}
\end{equation}
where energy scale of the singularity is given by
\begin{equation}
W_0=\sqrt{\frac{\bar{\delta} \Delta^2}{\Omega}}.
\label{W_0}
\end{equation}

Since $E_Z, \Delta \gg \bar{\delta}$,
one can neglect the fine structure of the DoS
on the scale of $\bar{\delta}$
and substitute the summation over $i$ by the integration over $\epsilon_i$:
\begin{eqnarray}
\sum_i G_{i\downarrow} &=& \nu_0 \int d\epsilon_i
 \frac{\omega_+ + \epsilon_i-E_Z/2-\Omega/2}
{-\epsilon_i^2 + (\omega_+ -E_Z/2-\Omega/2) -W^2_0 } \nonumber\\
&=& -i \nu_0 \pi \frac{\omega-E^*}{\sqrt{(\omega-E^*)^2- W^2_0} }.
\label{GFdown}
\end{eqnarray}

Analogously, the GF for the up-spin electron can be
obtained by changing the signs of $E_Z$ and $\Omega$, so
that the singularity occurs at $\omega= -E^*$.

Substituting Eq. (\ref{GFdown}) into Eq. (\ref{dos}),
we obtain the final expression for the tunneling DoS in the ultra-small
grains
\begin{equation}
\nu_{\uparrow (\downarrow)} (\omega)=
\nu_0 F_0\left(\frac{\omega \pm E^*}{W_0}\right), \quad
F_0(x)={\rm Re}\left(\frac{x^2}{x^2-1}\right)^{1/2}, 
\label{0d-dos}
\end{equation}
where $\nu_0$ is the bare DoS per one spin, energy $E^*$ is defined by
Eq.~(\ref{E^*}) and width of the singularity $W_0$ is given by
Eq.~(\ref{W_0}).  Equations (\ref{0d-dos}) and (\ref{W_0}) are the
main result of this section. They predict the hard gap in the spin
resolved density of states: $\nu_\sigma (\omega )$ vanishes at
$|\omega +\sigma E^*| < W_0$. Overall density of states
$\nu_\downarrow + \nu_\uparrow$ is suppressed by a factor of two near
the singularity.

In this calculation we neglected higher corrections to the self
energy, {\em e.g.}, those shown in Fig.~\ref{Fig:2.3}a, b.  
In order to justify this approximation, we have to compare
contributions shown in Figs.~\ref{Fig:2.3}a, b
with the reducible diagram shown in Fig.~\ref{Fig:2.3}c 
included in Eqs. (\ref{self}) and (\ref{GF}). 
Singular contribution originates from the pole of
$\Lambda$.  It means that $\Lambda$ carries frequency $\Omega$.  The
singularity in the DoS at $\omega=E^*$ appears when the pole of the
self-energy and the pole of $G^0$ coincide.  This happens when GF
$G^0$ for up-spin carries energy $\Omega-\omega$.  In diagrams
Fig.~\ref{Fig:2.3}a, b, and c, the intermediate $G^0$ for down-spin
should carry energy $\omega$ to give singularity to the DoS at $E^*$.
This condition can not be satisfied for the diagrams
Fig.~\ref{Fig:2.3}a, b.  As a result, after the integration over the
intermediate frequency, these higher order corrections turns out to be
smaller than the reducible contribution (c) by small factor
$W_0/\Delta\simeq \sqrt{\bar{\delta}/\Delta} \ll 1$.

\narrowtext
{\begin{figure}[ht]
\vspace{0.2cm}
\epsfxsize=7cm
\hspace*{0.5cm}
\epsfbox{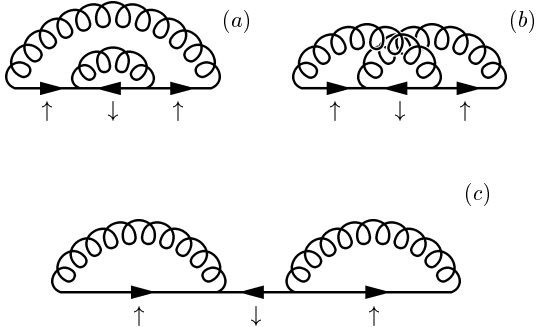}
\vspace{-0.05 cm}
\caption{
Higher order correction to the self-energy (a) and (b), which were
neglected in comparison with reducible diagram (c).
}
\label{Fig:2.3}
\end{figure} 
}
\widetext

\section{Disordered infinite systems}
\label{sec:hd}

In this section, we will obtain the quantitative results for the
tunneling anomaly in the infinite systems. In subsection \ref{sec:hdp} we
will start from the perturbation theory and demonstrate that the
lowest order perturbative results diverge algebraically at
energies close to $E^*$. In order to deal with this divergence, we
develop non-perturbative approach in subsection \ref{sec:hdn} and obtain
the analytic expressions for the shape of the singularities for all
interesting cases. This machinery will be also used later in
Sec.~\ref{sec:rp}.

\subsection{Perturbative results}
\label{sec:hdp}
The analysis of $0$-D system presented in the previous subsection, is
 not directly applicable to superconducting wires and films because
 one can not approximate interaction Hamiltonian
\begin{equation}
\hat{H}_{int}=- \lambda \nu_0^{-1} \int_{-\infty}^{\infty} dr
a^{\dagger}_{\uparrow}(r) a^{\dagger}_{\downarrow}(r)
a_{\downarrow}(r) a_{\uparrow}(r).
\label{eq:3.1.1}
\end{equation}
by its diagonal matrix elements.  Despite this complication, we will
still be able to show that the singularity persists and remains at
the same bias as that in $0$-D.

To describe this singularity we once again have to evaluate the effect
of the superconducting fluctuations on the GF of electrons.  First of
all, we need to evaluate the propagator of superconducting fluctuations
$\Lambda$, see Fig.~\ref{Fig:3.1}.  In contrast with $0$-D case, the
superconducting fluctuations in the bulk system are inhomogeneous: the
propagator for the superconducting fluctuations depends on the wave
vector $Q$.  (We will omit the vector notation in the momenta, e.g.,
$Q \equiv {\vec Q}$.)  
Solving the Dyson equation, shown in Fig.~\ref{Fig:3.1}a, we obtain the
propagator
\begin{equation}
\Lambda(\omega, Q)= \frac{2}{ \nu_0 {\rm ln} 
\left( \frac{E_Z^2-(|\omega|+i D Q^2)^2}
{\Delta^2} \right) },
\label{Lambda}
\label{Lamda}
\end{equation}
where $D$ is the diffusion coefficient. At $Q=0$, propagator
(\ref{Lambda}) resembles zero-dimensional expression
(\ref{Lam0}).  We see that the propagator has the singularity at
$\omega$ close to $\Omega$ provided $D Q^2 \ll \Omega$.  As we will see
shortly, it results in the singularity in the DoS developing at
exactly the same energy as in $0$-D case, $\omega=E^*$, given by
Eq.~(\ref{singular}).

{\narrowtext
\begin{figure}[ht]
\vspace{0.2 cm}
\epsfxsize=8cm
\centerline{\epsfbox{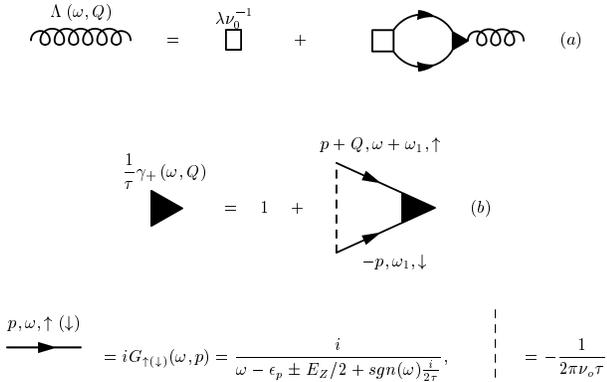}}
\vspace{0.1cm}
\caption{
Diagrams for the (a) propagator of superconducting fluctuations
$\Lambda \left(\omega, Q\right)$ and (b)
the vertex function $\gamma_+(\omega, Q)$.
}
\label{Fig:3.1}
\end{figure}
}
\widetext

Next step is to consider vertex function in the particle-particle
channel.  The ladder approximation which gives the main contribution
at $\epsilon_F\tau\gg 1$ ( $\tau$ is the elastic mean free time) is
shown in Fig.~\ref{Fig:3.1}b. Analytically, they are given by
\begin{equation}
\gamma_\pm (\omega,Q) 
= \tau  + I_\pm (\omega, Q) 
\gamma_\pm(\omega,Q).
\label{vertex}
\end{equation}
Here $I_\pm(\omega, Q)$ stands for
\begin{eqnarray}
I_\pm(\omega, Q) &=&\frac{1}{2\pi \nu_0 \tau} \int (dp) 
G_{\uparrow (\downarrow)}(\omega+\omega_1, \epsilon_{p})
G_{\downarrow (\uparrow )}(\omega_1,\epsilon_{-p-Q})\nonumber\\
&=& \frac{1}{2\pi \nu_0 \tau} \int (dp)
\frac{1}
{(\omega+\omega_1-\epsilon_p\pm \frac{E_Z}{2}
+\frac{i}{2\tau}sgn(\omega +\omega_1))
(\omega_1-\epsilon_{-p+Q}\mp \frac{E_Z}{2}
+\frac{i}{2\tau}sgn(\omega_1))}\nonumber\\
&=& 
\left\{
1+ \tau \left[i (|\omega|\pm sgn(\omega)E_Z) - D Q^2\right]
\right\}
\theta\left[ -\left(\omega+\omega_1\right)\omega\right],
\label{scatter}
\end{eqnarray}
where $(dp)=\frac{d^dp}{(2\pi)^d}$, and the GFs are averaged over
disorder\cite{AGD,AA}.  Here we used the conditions of the diffusion
approximation, $\omega \tau \ll 1$ and $Q l \ll 1$.
We substitute  Eq.~(\ref{scatter}) into Eq.~(\ref{vertex}) and we obtain
\begin{equation}
\gamma_\pm(\omega,Q) 
= \tau \theta\left[ \left(\omega+\omega_1\right)\omega\right]
+\frac{\theta\left[
-\left(\omega+\omega_1\right)\omega\right]}
{-i\left[|\omega| \pm sgn(\omega) E_Z\right] + DQ^2}.
\label{Gamma}
\end{equation}

{\narrowtext
\begin{figure}[ht]
\vspace{0.1 cm}
\hspace*{.25cm}
\epsfxsize=6cm
\epsfbox{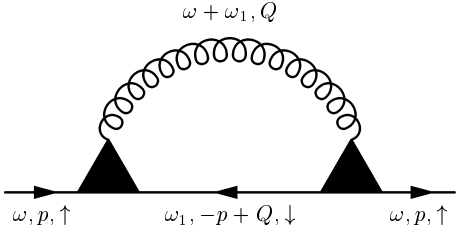}
\vspace{-0.4 cm}
\caption{
Diagrams for the lowest order
corrections to the one-particle Green function.
}
\label{Fig:3.2}
\end{figure}
}
\widetext

Now we can evaluate the first order
correction to the one-particle GF, Fig.~\ref{Fig:3.2}:
\begin{equation}
\delta G^{(1)}_{\downarrow}(\omega,p) = 
\frac{1}{\tau^2}\int (dQ) \int \frac{d\omega_1}{2\pi}
G^2_{\downarrow}(\omega, p) \Lambda(\omega+\omega_1, Q)
G_{\uparrow}(\omega_1,-p+Q)\nonumber\\
\gamma^2_{-} (\omega-\omega_1,Q)
\end{equation}
The DoS is determined by the GF integrated over $p$:
\begin{equation}
\nu_{\uparrow(\downarrow)}(\omega) =
-\frac{sgn(\omega)}{\pi} {\rm Im} 
 \int (dp) 
G_{\uparrow(\downarrow)}(\omega,p)
\label{firstorder}
\end{equation}

We substitute Eq.~(\ref{Gamma}) into Eq.~(\ref{firstorder}) and
perform the integration over $p$ and $\omega_1$.
For  $Ql\ll 1$ and $|\omega-\omega_1|\tau \ll 1$ integration over
the momentum $p$ results in
\[
\int (dp)G^2_{\downarrow}(\omega, p)G_{\uparrow}(\omega_1,-p+Q)
= i 2\pi \nu_0 \theta(-\omega\omega_1) \tau^2 sgn(\omega_1).
\]
Performing integration over $\omega_1$ we take into account only the
 pole contributions  in the propagator (\ref{Lambda}) for $DQ^2
\ll \Omega$:
\begin{equation}
\Lambda(\omega,Q) 
\approx \frac{\Delta^2}{\nu_0\Omega}
\left( \frac{1}{\Omega+ \omega - iDQ^2}
+ \frac{1}{\Omega - \omega - iDQ^2} \right).
\label{Lambdapole}
\end{equation}
since the integrals along the branch cuts give only the corrections
which are smooth  functions of $\omega$.
The main contribution to the frequency integral results from the
region where the real part of the pole of the propagator $\Lambda$ in
Eq.~(\ref{firstorder}), ${\rm Re}\, \omega_1 = -\omega \pm \Omega $ is
close to that of the vertex $\gamma_-$, ${\rm Re}\ \omega_1 = \omega +
sgn(\omega_1-\omega) E_Z$, and the imaginary parts of those poles have
different signs. The latter requires $\omega\,{\rm Re}\,\omega_1\, <0$, 
$\omega^2 > \left[{\rm Re}\,
\omega_1\right]^2 $. One can easily check that all these conditions
can be met only if $\omega$ is close to $E^*$ from Eq.~(\ref{E^*}).

Evaluating the integral over $\omega_1$ in Eq.~(\ref{firstorder}) we
 obtain the first order correction to the DoS 
\begin{equation}
\frac{\delta \nu^{(1)}_{\uparrow (\downarrow)}(\omega)}{\nu_0}
= - \frac{\Delta^2}{2 \nu_0 \Omega} {\rm Re} \int (dQ)
C^2 (\omega \pm E^*, Q),\label{1st-cooperon}
\end{equation}
where $C_{\uparrow (\downarrow)}(\omega, Q)$ is the Cooperon 
given by
\begin{equation}
C (\omega, Q) =
\frac{1}{-i |\omega| + D Q^2}.
\label{cooper}
\end{equation}
[Calculation of $\delta \nu_\uparrow$ requires an obvious modification of
Eq.~(\ref{firstorder}).]

For one dimensional case (wire), this correction acquires the form
\begin{equation}
\frac{\delta \nu^{(1)}_{\uparrow (\downarrow)}}{\nu_0}(\omega)
=  \frac{\sqrt{\Delta}}{8\nu_0 \Omega \sqrt{2 D} } \left(
\frac{\Delta}{|\omega \pm E^*|} \right)^{3/2}.\label{1st-1d-dos}
\end{equation}
It is possible to neglect higher order corrections to DoS
only provided $|\omega \pm E^*|$ is large.
For $\omega \rightarrow E^*$, correction  (\ref{1st-1d-dos})
diverges. Therefore we need to sum up all the orders of perturbation
theory to
describe the DoS in the vicinity of $E^*$.
Such calculation is carried out in the next subsection.

For two dimensions (films), the first order correction to the DoS
vanishes for $|\omega| \neq E^\ast$.  However, this is nothing but an
artifact of the first order approximation and  the second order
correction is already finite.  Diagrams for this correction are shown in
Fig.~\ref{Fig:3.3}.  The result can be written as:
\begin{equation}
\frac{\delta \nu^{(2)}(\omega)_{\uparrow (\downarrow)}}{\nu_0}
=-2 \left(\frac{\Delta^2}{4\nu_0\Omega}\right)^2 
\frac{\partial}{\partial\omega}
{\rm Im} \int (dQ_1)(dQ_2) 
C^2_{\uparrow(\downarrow)}(\omega, Q_1)C_{\uparrow(\downarrow)}(\omega, Q_2) .
\label{2nd-DOS}
\end{equation}
For two dimensional case Eq.~(\ref{2nd-DOS})
gives
\begin{equation}
\frac{\delta \nu^{(2)(\omega)}_{\uparrow (\downarrow)}}{\nu_0}
=- \left[\frac{\Delta^2}{4 g \Omega (\omega \pm E^*)}\right]^2 
{\rm ln} \left( \frac{\Omega}{|\omega \pm E^*|} \right)^2 ,
\label{2nd-2d-dos}
\end{equation}
where $g=4 \pi D \nu_0 \gg 1$ is the dimensionless conductance of the
film in the normal state.  Deriving Eq.~(\ref{2nd-2d-dos}), we cut off the
logarithmic divergence at large momenta $Q_2$ by the condition $DQ^2_2
\lesssim \Delta$, since it determines the applicability of a single
pole approximation (\ref{Lambdapole}).

{\widetext
\begin{figure}[ht]
\vspace{-0.2 cm}
\epsfxsize=14cm
\centerline{
\epsfbox{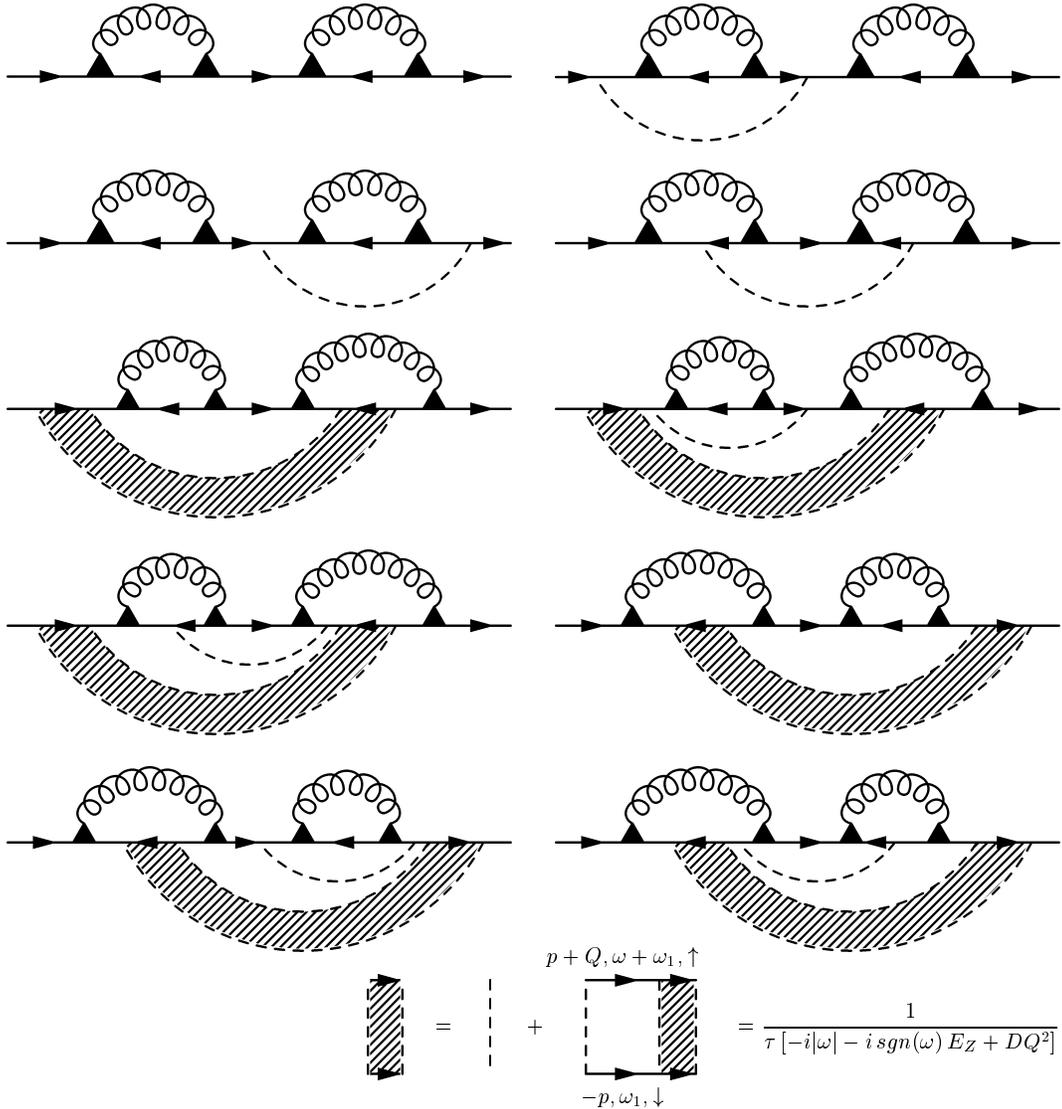}
}
\vspace{0.3cm}
\caption{
Diagrams for the second order corrections to the DoS.
Diagrams irreducible with respect to the curly lines, similar
to Fig.~\protect\ref{Fig:2.3}a,b, are negligible for the reason
discussed in the end of Sec.~\protect\ref{sec:0d}.
}
\vspace{-0.26cm}
\label{Fig:3.3}
\end{figure}
}
\widetext

 As well as $1D$ case, the perturbation theory fails to describe the
DoS in the vicinity of $E^*$ in two dimensions.  It is noteworthy that
the singularity described by Eq.~(\ref{2nd-2d-dos}) is much more
pronounced than that of the normal metal $(T > T_c)$ which arises due
to the superconducting fluctuations and it is of the order of $g^{-1}
{\rm ln} [{\rm ln}(\omega \pm E_Z)]$, see Ref.~\onlinecite{AA}.
The enhancement of this singularity results directly from the isolated
pole in the propagator of the superconducting fluctuations, see Eqs.~
(\ref{Lambda}) and (\ref{Lambdapole}) rather than in the branch cuts
of Ref.~\onlinecite{AA}.

\subsection{Non-perturbative results}
\label{sec:hdn}

\subsubsection{Derivation of self-consistency equations}
We start summation of the perturbation theory terms from the series of
diagrams for DoS presented on Fig.~\ref{Fig:3.4}.  Such diagrams
dominate in each order of the perturbation theory in comparison with
those of the same order but including irreducible with respect to
curly lines elements, similar to Fig.~\ref{Fig:2.3}a,b. The later
statement can be justified by arguments similar to the analysis of
diagrams Fig.~\ref{Fig:3.4} presented in the end of Sec.~\ref{sec:0d}.

{\narrowtext
\begin{figure}[ht]
\vspace{0.5 cm}
\hspace*{.25cm}
\epsfxsize=7cm
\epsfbox{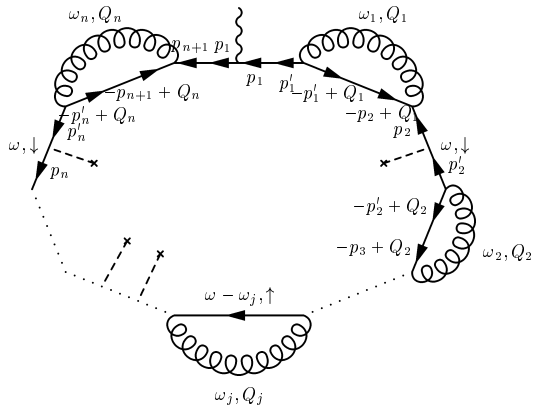}
\vspace{0.1cm}
\caption{ Structure of the $k$-th order perturbation theory.  Vertical
wiggly line on the upper Green function corresponds to cutting this GF
into two lines and fixing its frequency $\omega$.  }
\vspace{0.02cm}
\label{Fig:3.4}
\end{figure}
}
\widetext

The diagram of the k-th order of the perturbation theory,
Fig.~\ref{Fig:3.4}, contains $k$ curly lines that stand for the
fluctuation propagator $\Lambda(\omega_j,Q_j)$ in Eq~ (\ref{Lamda})
($1 \leq j \leq k$).  It also contains $(2k+1)$ GF.  Before averaging
over disorder, each GF $G_{\sigma}(\omega;p,p^{\prime})$ depends on
two momenta (initial $p$ and final $p^{\prime}$) and on the direction
of the spins $\sigma = \pm \equiv \uparrow, \downarrow$.  Contribution
of this diagram to the DoS of $\downarrow$ electrons $\delta
\nu_{\downarrow}(\omega)$ at positive energy $\omega > 0$ is
\begin{eqnarray}
\delta \nu_{\downarrow}^{(k)}(\omega) &=& 
-\frac{sgn(\omega)}{\pi} {\rm Im}
\int \frac{d\omega_1 d\omega_2 \cdots 
d\omega_k}{(2\pi)^k}
\int (dp_1)(dp_1^{\prime}) \cdots (dp_{k+1})
\int (dQ_1) (dQ_2) \cdots (dQ_k) \nonumber\\
&\times&
G_{\downarrow}(\omega;p_{k+1}, p_1)
 \prod_{j=1}^k \Lambda(\omega_j,Q_j) 
G_{\downarrow}(\omega;p_j,p_j^{\prime}) 
G_{\uparrow}(-\omega+\omega_j;-p_j^{\prime}+Q_j, -p_{j+1}+Q_j).
\label{nth}
\end{eqnarray}
Disorder averaging of the product of the Green functions in
Eq.~(\ref{nth}) and of the superconducting propagator $\Lambda$ can be
carried out independently, since their correlation gives rise to
non-singular corrections containing additional smallness
$1/(\epsilon_F\tau)$.

For the sake of convenience we introduce
$G^{R(A)}(\omega;p,p^{\prime})$: retarded (advanced) GF at $E_Z=0$.
In the absence of spin-orbit scattering the Green functions
$G^{R(A)}(\omega;p,p^{\prime})$ from Eq.~(\ref{nth}) can be presented
through $G^{R(A)}$ as
\begin{equation}
G_{\sigma}(\omega;p,p^{\prime}) = \left\{
\begin{array}{ll}
G^R(\omega+\sigma E_Z/2;p,p^{\prime}) & \mbox{$\omega > 0$}\\
G^A(\omega+\sigma E_Z/2;p,p^{\prime}) & \mbox{$\omega < 0$}
\end{array}
\right.
\label{RA}
\end{equation}

After substituting Eqs. (\ref{RA}) and (\ref{Lambdapole}) into
Eq.~(\ref{nth}), one can integrate over all intermediate frequencies 
$\omega_j$, because the fluctuation propagator $\Lambda(\omega_j, Q_j)$
has simple poles at $\omega_j=\pm \Omega \mp i DQ^2_j$.
According to Eq.~(\ref{RA}), $G_{\downarrow}(\omega;p,p^{\prime})$ in Eq.~
(\ref{nth}) is retarded ($\omega > 0$).
In order to get the advanced GF from $G_{\uparrow}(\omega_j-\omega;
-p_j^{\prime}+Q_j,-p_{j+1}+Q_j)$ we choose the pole in
$\Lambda(\omega_j;Q_j)$ with positive real part,
$\omega_j=\Omega-iDQ^2_j$.
(One can check that another pole leads to the product of the retarded
functions, which vanish after the disorder averaging.)
Using Eqs.~(\ref{Lambdapole}) and (\ref{omega}) and introducing
short-hand notation 
\begin{equation}
\omega_{Q_j}
\equiv -\omega+2E^*-iDQ^2_j,
\label{OmQ}
\end{equation}
we can present $\delta \nu_{\downarrow}(\omega)$
at $\omega > 0$ as 
\begin{eqnarray}
\delta \nu_{\downarrow}^{(k)}(\omega) &=& 
-\frac{1}{\pi} \left(\frac{\Delta^2}{\nu_0 \Omega}\right)^k
{\rm Im} \int (dp_{k+1}) G^R(\omega; p_{k+1},p_1) \prod_{j=1}^n(dp_j)
 (dp_j^{\prime}) (dQ_j) \nonumber\\
& & G^R(\omega;p_j,p_j^{\prime}) 
G^A(\omega_{Q_j};-p_j^{\prime}+Q_1,-p_{j+1}+Q_j).
\label{nthnu}
\end{eqnarray}
Averaging  Eq. (\ref{nthnu}) over disorder and using the identity
\begin{equation}
\int(dp_1) G^R(\omega;p_1,p_1^{\prime})
G^R(\omega;p_{k+1},p_1)
=-\frac{\partial}{\partial \omega}
G^R(\omega;p_{k+1},p_1^{\prime}),
\end{equation}
we obtain
\begin{eqnarray}
\langle \delta \nu_{\downarrow}^{(k)}(\omega) \rangle &=& -
\frac{1}{\pi}
\left( \frac{\Delta^2}{\nu_0 \Omega} \right)^k
\left(\frac{1}{2k}\right) \frac{\partial}{\partial  \omega}
{\rm Im}  \langle\int \prod_{j=1}^k (dQ_j) (dp_{j}) 
(dp_{j}^{\prime})\nonumber\\
& \times  & 
G^R(\omega;p_{j}, p_{j}^{\prime}) 
G^A(\omega_{Q_j};-p_j^{\prime}+Q_j,-p_{j+1}+Q_j)
\rangle. 
\label{nthdos1}
\end{eqnarray}
The role of the factor $1/k$ in Eq.~(\ref{nthdos1}) is to cancel the
multiple counting: there are $k$ retarded GF in Eq. (\ref{nthdos1}),
and application of the operator $\partial_\omega$ to any one of them
leads to Eq. (\ref{nthnu}).  In addition, Eq. (\ref{nthdos1}) includes
terms like $\partial G^A/\partial \omega$.  One can check that they
give exactly the same contribution as terms which contain $\partial
G^R/\partial \omega$.  Additional factor $1/2$ takes these
contributions into account.

Using a conventional trick $1/k=\int_0^1
\eta^k d\eta$, (see Ref.~\onlinecite{AGD}), one can present 
$\langle \delta \nu^{(k)}_{\downarrow}(\omega)\rangle $
in a form
\begin{eqnarray} 
\langle \delta\nu^{(k)}_{\downarrow}(\omega) \rangle &=&
-\frac{1}{2\pi\tau} \frac{\Delta^2}{\nu_0 \Omega}
{\rm Im} \frac{\partial}{\partial \omega}
\int^1_0 d\eta \int (dQ) (dp) \langle G^R(\omega;p) \rangle
\langle G^A(\omega_Q;-p+Q) \rangle \nonumber\\
& \times & \gamma^{(k)}_\omega (\tilde{\omega}_{Q}, Q),
\label{nthdos2}
\end{eqnarray}
where  short hand notation
\begin{equation}
\tilde{\omega}_{Q}=\omega-\omega_Q=2\omega-2 E^*+i DQ^2
\label{omegaq}
\end{equation}
is used and $\omega_Q$ is defined in Eq.~(\ref{OmQ}).
Similar equation holds for spin up and negative $\omega$.
In this case one should use
$\tilde{\omega}_{Q}=-2\omega-2E^*+iDQ^2$.

The vertex 
$\gamma^{(k)}$ can be written using Fig.~\ref{Fig:3.5}.
The rules of reading   diagrams on Figs.~\ref{Fig:3.5} and
\ref{Fig:3.100} are slightly different from the conventional rules we
used before: (i) to each curly line corresponds factor
$\eta \Delta^2/\left(\nu_0\Omega\right)$; (ii) no summation over the
frequencies is implied: each retarded GF bears frequency $\omega$,
each advanced GF bears frequency $\omega_Q$ given by Eq.~(\ref{OmQ});
(iii) each interaction with curly lines changes retarded Green
function to advanced and back.
The resulting expression reads:
\begin{eqnarray}
\gamma^{(k)}_\omega (\tilde{\omega}_Q,Q) & = &
\tau 
\left(\eta \frac{\Delta^2}{\nu_0 \Omega} \right)^{k-1}
\langle \int \prod_{j=1}^{k-1} (dQ_j) 
(dp_{j}) (dp_{j}^{\prime})\nonumber\\
 & & G^R(\omega;p_{j}, p_{j}^{\prime}) 
G^A(\omega_{Q_j};-p_j^{\prime}+Q_j,-p_{j+1}+Q_j) \nonumber\\
& \times  & G^R(\omega; p_1^{\prime},p) 
G^A(\omega_Q;-p_k^{\prime}+Q,-p+Q)
\rangle.
\label{gamman}
\end{eqnarray} 
Note that the averaged product of GFs in the integrand of 
Eq.~(\ref{gamman}) does not depend on $p$. 
We can thus perform in Eq. (\ref{nthdos2}) the integration over $p$.
As a result, Eq.~(\ref{gamman}) takes a form
\begin{equation}
\frac{\delta\nu}{\nu_0} = -\frac{\Delta^2}{\nu_0 \Omega}
{\rm Im} \frac{\partial}{\partial \omega} \int_0^1 d\eta
\int (dQ)  \gamma (\tilde{\omega}_Q,Q),
\label{corrdos}
\end{equation}
where 
\begin{equation}
\gamma(\tilde{\omega}_Q,Q)=\sum_{k=1}^{\infty} 
\gamma^{(k)}(\tilde{\omega}_Q,Q).
\end{equation}

{\narrowtext
\begin{figure}[ht]
\vspace{0.5 cm}
\hspace*{0.25cm}
\epsfxsize=7cm
\epsfbox{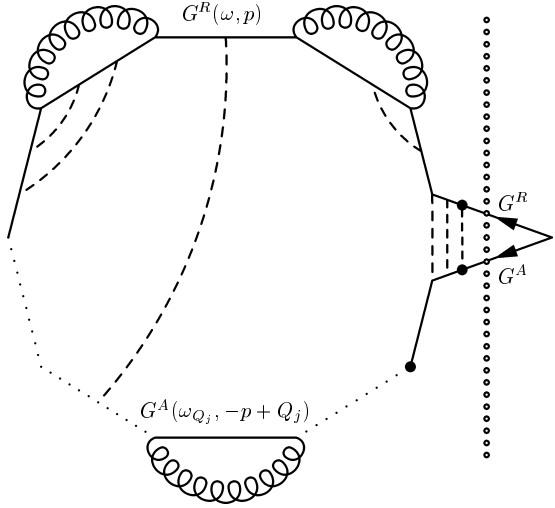}
\vspace{0.1cm}
\caption{
Structure  of the $\gamma^{(k)}$. Three points defining vertex are
denoted by thick dots.
}
\vspace{0.02cm}
\label{Fig:3.5}
\end{figure}
}
\widetext

Our goal is to evaluate $\gamma(\tilde{\omega}_Q,Q)$ self-consistently.
The difficulty is that diagrammatic series for $\gamma$
contains other elements except $\gamma$ itself, see
Fig.~\ref{Fig:3.100}.
{\narrowtext
\begin{figure}[ht]
\vspace{0.5 cm}
\epsfxsize=7cm
\centerline{\epsfbox{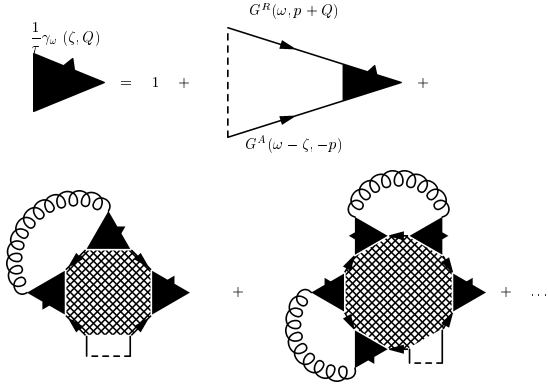}}
\vspace{0.5cm}
\caption{
Diagrammatic equation for the vertex $\gamma_\omega(\zeta,Q)$. 
}
\vspace{0.02cm}
\label{Fig:3.100}
\end{figure}
}
\widetext

These elements are known as Hikami boxes\cite{Hikami}.
The simplest Hikami box $B^{(2)}$, which appears already in
second order of the perturbation theory for $\delta \nu$, is
the integral $(dp)$ of a sum of three diagrams shown on Fig.~\ref{Fig:3.6}
\begin{equation}
B^{(2)}(\omega_1,\omega_2,Q_1,Q_2) 
=  \frac{1}{(2\pi \nu_0)^3}
[-i(\omega_1+\omega_2)+D(Q_1^2+Q_2^2)].
\label{hikami2}
\end{equation}

{\narrowtext
\begin{figure}[ht]
\vspace{0.2 cm}
\epsfxsize=7.5cm
\centerline{
\epsfbox{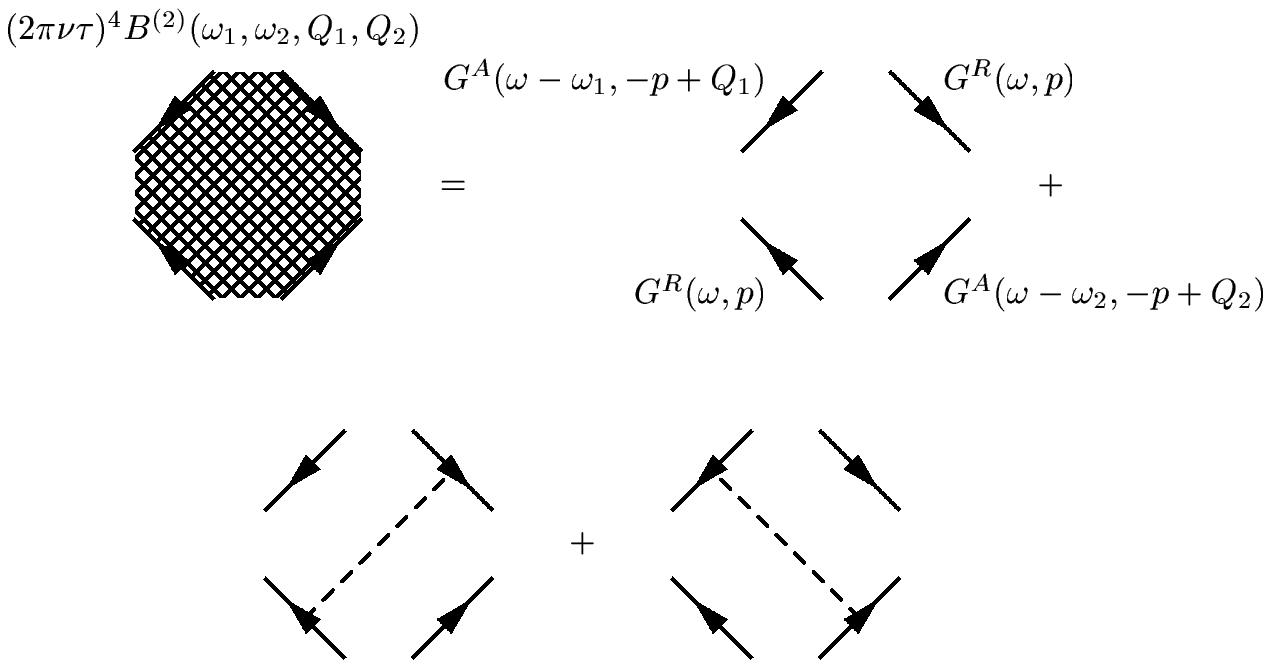}
}
\vspace{0.4cm}
\caption{
Second order Hikami box $B^{(2)}$.
}
\vspace{-0.02cm}
\label{Fig:3.6}
\end{figure}
}
\widetext

The third order Hikami box, $B^{(3)}$, is given by the momentum integral 
$(dp)$ of a sum of sixteen diagrams of Fig.~\ref{Fig:3.7}:
\begin{equation} 
B^{(3)}(\omega_1,\omega_2,\omega_3;Q_1,Q_2,Q_3) = 
-\frac{2}{(2\pi \nu_0)^5} [-i(\omega_1+\omega_2+\omega_3)
+D(Q_1^2+Q_2^2+Q_3^2)].  
\label{hikami3}
\end{equation}

{\narrowtext
\begin{figure}[ht]
\vspace{0.2 cm}
\epsfxsize=7.5cm
\centerline{
\epsfbox{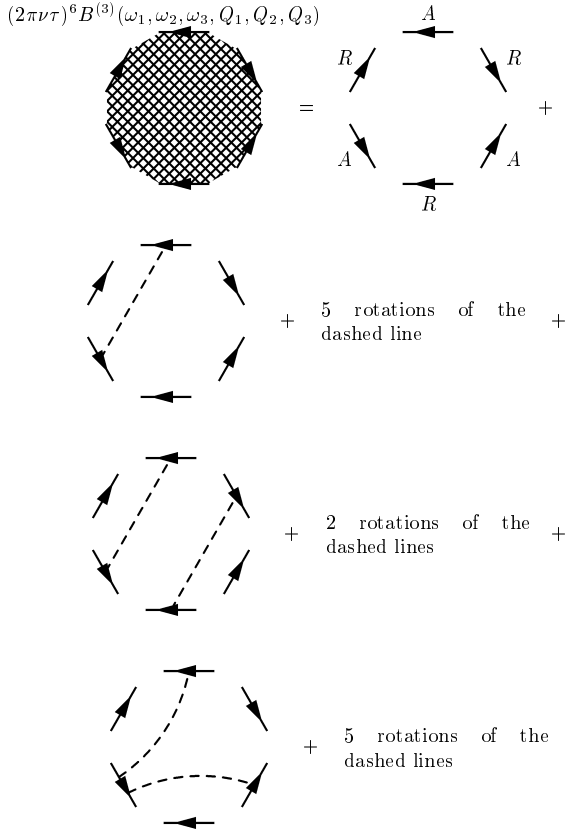}
}
\vspace{0.4cm}
\caption{
Third order Hikami box $B^{(3)}$. Momenta and frequencies in the Green
functions are arranged similar to Fig.~\protect\ref{Fig:3.6}.
}
\vspace{-0.02cm}
\label{Fig:3.7}
\end{figure}
}
\widetext

The equation for the vertex $\gamma(\tilde{\omega}_Q,Q) \equiv
\gamma_{\omega}(\zeta=\tilde{\omega}_Q,Q)$ 
that determines correction to the DoS [Eq. (\ref{corrdos})]:
\begin{eqnarray}
\frac{1}{\tau} \gamma_{\omega}(\zeta, Q) & =&  1+ 
\left(\frac{1}{\tau}+i \zeta- D Q^2\right)
 \gamma_{\omega}(\zeta, Q)\nonumber\\\
& +& \eta \left(\frac{\Delta^2}{\nu_0 \Omega}\right) 
\int (dQ_1) (2\pi \nu_0)^3 B^{(2)}(\tilde{\omega}_Q,\tilde{\omega}_{Q_1},Q,Q_1)
\gamma^2(1) \gamma_{\omega}(\zeta,Q)\nonumber\\
& +& \eta^2 \left(\frac{\Delta^2}{\nu_0 \Omega}\right)^2 
\int \int (dQ_1) (dQ_2)(2\pi \nu_0)^5 
 B^{(3)}(\tilde{\omega}_Q,\tilde{\omega}_{Q_1}
,\tilde{\omega}_{Q_2},Q,Q_1,Q_2)\nonumber\\
&\times & \gamma^2(1) \gamma^2(2)
\gamma_{\omega}(\zeta,Q)+  \cdots,
\label{vertex1}
\end{eqnarray}
where $\gamma(j)=\gamma_{\omega}(\tilde{\omega}_{Q_j},Q_j)$
with $\tilde{\omega}_{Q_j}$ determined by Eq. (\ref{omegaq}).
We introduced extra variable $\zeta$ in order to separate the external
energy $\omega$ and the integration variable in Eq. (\ref{vertex1})
and further.

Hikami box of the k-th order $B^{(k)}$ is given by a sum of diagrams
(of the type of Figs.~\ref{Fig:3.6} and \ref{Fig:3.7}) which contain
$2k$ vertices.  Strictly speaking, $B^{(k)}$ depends on $2k-1$ sets of
momentum and energy transfer $(\omega,Q)$.  However, in order to
evaluate the DoS correction given by diagrams Fig.~\ref{Fig:3.4}, we
can restrict ourselves by Hikami boxes that depend only on $k$ sets
$(\omega_j,Q_j)$, $ 1\leq j \leq k$: $(\omega_j,Q_j)$ and
$(-\omega_j,-Q_j)$ characterize neighboring vertices.  Equations
(\ref{hikami2}) and (\ref{hikami3}) allow us to conjecture that
$B^{(k)}$ at arbitrary $k$ has a form
\begin{equation}
B^{(k)}\{\omega_j,Q_j\} =\frac{C_k}{(2\pi \nu_0)^{2k-1}}
\sum^k_{j=1} (-i\omega_j+DQ^2_j),
\label{hikamin}
\end{equation} 
where $C_k$ are numerical coefficients.

We are not going to determine coefficients $C_k$ directly.  An
important feature of Hikami boxes is that they are local objects,
determined by distances smaller or of the order of the mean free path
$l$.  Therefore, the coefficients $C_k$ in Eq.~(\ref{hikamin}) do not
depend on the dimensionality. Therefore, we can compare the exact
solution of $0$-D problem, Sec.~\ref{sec:0d}, with the sum of
perturbation theory series involving coefficients $C_k$ and thus find
those coefficients.

Equation (\ref{vertex1})  can be rewritten in terms of $C_k$ as
\begin{eqnarray}
(-i\zeta+DQ^2)\gamma_{\omega}(\zeta,Q) &=& 
1+(-i\zeta+DQ^2)\gamma_{\omega}(\zeta,Q) \sum^{\infty}_{k=1}
C_k\left[\eta \frac{\Delta^2}{\nu_0 \Omega} \int (dQ_1)
\gamma(1)^2\right]^k 
\\
&+& \gamma_{\omega}(\zeta,Q) \sum^{\infty}_{k=1}
 kC_k\left[\eta \frac{\Delta^2}{\nu_0\Omega}
\int (dQ_1) \gamma_{\omega}(\tilde{\omega}_{Q_1},Q_1)^2\right]^{k-1} 
\int (dQ_2) (-i\tilde{\omega}_{Q_2}+DQ_2^2) 
 \gamma_{\omega}(\tilde{\omega}_{Q_2},Q_2)^2.
\nonumber
\end{eqnarray}

Introducing the function, $f(x)$
\begin{equation}
f(x) =-\sum_{k=1}^{\infty} C_k x^k+1, 
\label{fx}
\end{equation}
we obtain a simple equation for $\gamma_{\omega}(\zeta,Q)$
\begin{equation}
(-i\zeta + DQ^2) \gamma_{\omega}(\zeta,Q) f(\beta_0) +
\gamma_{\omega}(\zeta,Q) \beta_1 f^{\prime} (\beta_0) = 1,
\label{Eq-vertex}
\end{equation}
where
$\beta_0$ and $\beta_1$ are connected with $\gamma$ by 
\begin{eqnarray}
\beta_0(\omega) &=& \eta \frac{\Delta}{\nu_0 \Omega} \int (dQ)
\gamma^2_{\omega}(\tilde{\omega}_Q,Q),\nonumber\\
\beta_1(\omega) &=& \eta \frac{\Delta}{\nu_0 \Omega} \int (dQ)
\gamma^2_{\omega}(\tilde{\omega}_Q,Q)
(-i\tilde{\omega}_Q+DQ^2).\label{betas}
\end{eqnarray}

One can use Eq. (\ref{Eq-vertex}) to present $\gamma_{\omega} (\zeta,Q)$
in the form 
\begin{equation}
\gamma_{\omega}(\zeta,Q) = \frac{Z(\omega)}{-i\zeta+DQ^2+2m(\omega)},
\label{gammaz}
\end{equation}
where parameters $Z$ and $m$ are determined as
\begin{equation}
Z(\omega)=\frac{1}{f(\beta_0(\omega))}, \;\;\; 
2 m(\omega)=\beta_1(\omega)\frac{f^{\prime}(\beta_0(\omega))}
{ f(\beta_0(\omega))}.
\label{zm}
\end{equation}

It was already mentioned that the Hikami boxes $B^{(n)}$ and thus
coefficients $C_n$ as well as the function $f(x)$ do not depend on the
number of dimensions $d$. 
In contrast, $\omega$-dependence of parameters $\beta_0$, $\beta_1$ and
$m$ are different at different dimensions. 
Let us first consider $0$D case in order to determine $f(x)$
explicitly.
At $d=0$ one has to abolish integration over $Q$ in Eq. (\ref{betas})
and substitute inverse mean level spacing $\bar{\delta}^{-1}$ for
$\nu_0$ and $2(\omega-E^*)$ for $\tilde{\omega}_Q$.
The vertex $\gamma$ can be determined straightforwardly
\begin{equation}
\gamma_{\omega}(\zeta)=\int\frac{d\epsilon_i}{2\pi} G_{i\downarrow}
(\omega)G_{i\uparrow}^{0*}(\omega-E_Z-\zeta),
\label{gammaw}
\end{equation}
where 0D GF $G_{i\sigma}$ and $G_{i\sigma}^0$ are determined by
Eq.~(\ref{GF}) [Eq.~(\ref{W_0}) for $W_0$ should be multiplied by
$\sqrt{\eta}$] and by Eq.~(\ref{G_0}) respectively.
Substitution of Eqs.~(\ref{GF}) and (\ref{G_0}) into Eq.~(\ref{gammaw})
gives after the integration over $\epsilon_i$
\begin{equation}
\gamma_{\omega}(\zeta)=\frac{1}{2}\left[F_0\left(\frac{\omega-E^*}{W_0}\right)+1\right]
\frac{1}{-i \zeta+i(\omega-E^*)[1-1/F_0(\frac{\omega-E^*}{W_0})]},
\label{gammaw2}
\end{equation}
where the function $F_0(x)$ is given by Eq. (\ref{0d-dos}),
$E^*$ is determined by Eq. (\ref{E^*}), and $W_0^2=\eta \bar{\delta}
\Delta^2/\Omega$.
By comparing Eq.~(\ref{gammaw2}) with Eq.~(\ref{gammaz}) we immediately
obtain $Z(\omega)$ and $m(\omega)$ for zero dimensions:
\begin{eqnarray}
Z_0(\omega)&  =& \frac{F_0(\frac{\omega-E^*}{W_0})+1}{2},\label{z0w}\\
m_0(\omega)&  =& i (\omega-E^*)
\frac{F_0(\frac{\omega-E^*}{W_0})-1}{2F_0(\frac{\omega-E^*}{W_0})}.
\label{m0w}
\end{eqnarray}
On the other hand, from zero dimensional version of Eq.~(\ref{betas})  
at $\zeta=2\omega-2E^*$ we can determine $\beta_0$ and $\beta_1$:
\begin{equation}
\beta_0 = \frac{1}{4 F_0^2}, \;\;\; \beta_1=-2i\omega =\frac{-i\omega}{F_0^2}.
\end{equation}

Now we are in state to determine the function $f(x)$ from Eq.~(\ref{fx}).
We express $F_0$ through $\beta_0$, substitute it into Eq.~(\ref{z0w})
for $Z_0$ and use the connection Eq. (\ref{zm}) between $Z_0(\omega)$
and $f(\beta_0)$.
As a result we have
\begin{equation}
f(x)=\frac{1}{2x}-\frac{1}{F_0(2x)}=
\frac{1}{2x}-\frac{\sqrt{1-4x}}{2x}.
\label{fucf}
\end{equation}
This functional dependence which remains the same at all dimensions, can
be used to evaluate $\delta\nu_{\downarrow}(\omega)$ for $d=1,2$.

Equations (\ref{corrdos}), (\ref{betas}) -- (\ref{zm}) and
(\ref{fucf}) constitute complete set  allowing to find the DoS in
any dimensions. 
One has to substitute Eq.~(\ref{gammaz}) into Eq.~(\ref{betas}),
and find functions $Z(\omega)$ and $m(\omega)$ self-consistently
with the help of Eqs.~(\ref{zm}) and (\ref{fucf}).
The result should be substituted in Eq.~(\ref{corrdos}) which gives
the final non-perturbative answer for the DoS.
In the following subsection, this program will be carried out for 1D
(wires) and 2D (films) systems.

\subsubsection{Solution of self-consistency equations}
\label{sec:slce}

We substitute Eq.~(\ref{gammaz}) into Eqs.~(\ref{betas}) and
(\ref{corrdos}) and perform integration over the wavevectors $Q$.
Equation (\ref{corrdos}) acquires the form 
\begin{equation}
\frac{\delta \nu_\sigma}{\nu_0} = - 2 \frac{\partial}{\partial\omega}
{\rm Im}\int_0^1 d\eta \frac{M_d\left[-i\omega+m
(\omega,\eta)\right]}{f\left[\beta_0(\omega,\eta)\right]}.
\label{eq:3.3.1}
\end{equation}
Here, we used Eq.~(\ref{zm}) and introduced dimensionless frequency
and mass
\begin{equation}
\omega \to \frac{\omega + \sigma E^*}{W_d}, \quad 
m \to \frac{m}{W_d}.
\label{eq:3.3.2}
\end{equation}
The relevant energy scales, which, as we will see below, are the widths of the
tunneling anomaly, are given by
\begin{equation}
W_1 = 3\left(\frac{ \Delta^2}
{16 \nu_1 \Omega \sqrt{D} } \right)^{2/3},
\quad
W_2 = \frac{\Delta^2}{4g \Omega},
\label{eq:3.3.3}
\label{Wd}
\end{equation}
where $\Omega$ is given by Eq.~(\ref{omega}), $D$ is the diffusion
coefficient, $\nu_1 = (mp_FS)/(2\pi^2)$ is the one dimensional
density of states per unit spin ($S$ is the cross-section of the
wire), and $g = 4\pi\nu_2 D$ is the dimensionless
conductance\cite{footnote}. 
The latter is related to the normal state resistance of the film as
$g=25.4k\Omega/R_\Box$.

Dimensionless functions in Eq.~(\ref{eq:3.3.1}), $M_d(x)$, are defined
as
\begin{equation}
M_1(x) = \frac{2}{\sqrt{ 3 x}}, \quad M_2(x)= \ln\left(\frac{4 g}{x}\right).
\label{Md}
\label{eq:3.3.4}
\end{equation}
To find $M_2$  we cut off the logarithmic
divergence at large momenta $Q$ by the condition $DQ^2 \lesssim
\Delta$, which determines the applicability of a single pole
approximation (\ref{Lambdapole}), and neglected the factor
$\Delta/\Omega \simeq 1$ in the argument of the logarithm.  

\begin{mathletters}
\label{eq:3.3.5}
Using the same notation, we obtain from Eqs.~(\ref{betas}) and (\ref{zm})
\begin{eqnarray}
&
\beta_0(\omega, \eta)\left\{ f[\beta_0(\omega, \eta)]\right\}^2 =
- \eta M_d^\prime\left[-i\omega+m
(\omega,\eta)\right]& 
\label{eq:3.3.5a}\\
&m\left\{ f[\beta_0(\omega, \eta)]\right\}^3  =
\eta f^\prime[\beta_0(\omega, \eta)]
\left\{
 M_d \left[-i\omega+m(\omega,\eta)\right] +
m (\omega,\eta) 
M_d^\prime\left[-i\omega+m (\omega,\eta)\right]
\right\}.&
\label{eq:3.3.5b}
\end{eqnarray}
\end{mathletters}
\begin{mathletters}
\label{eq:3.3.6}
Equation~(\ref{fucf}) allows to formally solve Eq.~(\ref{eq:3.3.5a}):
\begin{eqnarray}
&
\displaystyle{
\beta_0=-\frac{\eta M_d^\prime\left(-i\omega+m\right)}
{\left[1- \eta M_d^\prime\left(-i\omega+m\right)\right]^2},
}&
\label{eq:3.3.6a}\\
&
\displaystyle{
f(\beta_0) = 1- \eta M_d^\prime\left(-i\omega+m\right),
}&
\label{eq:3.3.6b}\\
&
\displaystyle{
\frac{ f^\prime (\beta_0)}{\left[ f(\beta_0)\right]^3}=
\frac{1}{ 1 + \eta M_d^\prime\left(-i\omega+m\right)}.
}&
\label{eq:3.3.6c}
\end{eqnarray}
\end{mathletters}
We can now substitute Eq.~(\ref{eq:3.3.6c}) into
Eq.~(\ref{eq:3.3.5b}) and obtain after simple algebra
\begin{equation}
m(\omega,\eta) =\eta  
M_d \left[-i\omega+m(\omega,\eta)\right].
\label{eq:3.3.7}
\end{equation}

Further calculation are substantially simplified due to the fact that
the integrand in Eq.~(\ref{eq:3.3.1}) can be presented as a total
derivative with respect to $\eta$. In order to demonstrate this, we
differentiate both sides of Eq.~(\ref{eq:3.3.7}) with respect to
$\eta$:
\begin{equation}
\frac{\partial m}{\partial \eta } =  
M_d \left(-i\omega+m \right)
+ \eta M_d^\prime \left(-i\omega+m \right)\frac{\partial m}{\partial \eta }.
\label{eq:3.3.8}
\end{equation}
Finding $\partial m/\partial\eta$ from Eq.~(\ref{eq:3.3.8}), we notice
with the help of Eq.~(\ref{eq:3.3.6b}) that it coincides with the integrand
in Eq.~(\ref{eq:3.3.1}).  Integration in Eq.~(\ref{eq:3.3.1}) can be
immediately performed and we obtain
\begin{equation}
\frac{\delta \nu_\sigma}{\nu_0} = - 2 \frac{\partial}{\partial\omega}
{\rm Im}\  m(\omega,\eta=1).
\label{eq:3.3.9}
\end{equation}

Finally we put $\eta=1$ in Eq.~(\ref{eq:3.3.7}), differentiate both
sides of this equation with respect to $\omega$, and substitute the result
for $\partial m/\partial\omega$ into Eq.~(\ref{eq:3.3.9}). 
After restoration of original
units for $\omega$ according to Eq.~ (\ref{eq:3.3.2}),  we obtain for the
density of states $\nu_\sigma (\omega) = \nu_0 +
\delta\nu_\sigma (\omega)$ the following result
\begin{equation}
\frac{\nu_\sigma (\omega)}{\nu_0} = F_d \left(\frac{\omega +
\sigma E^*}{W_d}  \right)
\label{DoSfinal}
\end{equation}
where $\sigma =\pm 1$ corresponds to the spin-up and spin-down
densities of states respectively, and the widths of the singularity $W_d$
are defined in Eq.~(\ref{Wd}). Dimensionless function $F_d(x)$ is given
by
\begin{mathletters}
\label{Fd}
\begin{equation}
F_d(x) = {\rm Re}\frac{1 + z(x)}{1- z(x)}
\label{F}
\end{equation}
where function $z(x)$ is implicitly defined as the solution of
equations 
\begin{eqnarray}
&z(x) = M^\prime_d\left[ -ix +y(x) \right],&\nonumber
\\
&y(x) = M_d\left[ -ix + y(x) \right],&
\label{y}
\end{eqnarray}
with functions $M_d(x)$ being defined in Eq.~(\ref{Md}).
If Eq.~(\ref{y}) has several solutions, one has to choose the one
reproducing the perturbation theory at $x \gg 1$ and remaining on the
same list of the Riemann surface at small $x$.
\end{mathletters}

In 1D case, Eq.~(\ref{y}) can be rewritten as a cubic equation and
solved using the Cardano formula which yields universal 
(independent on $\nu_1$ and $D$) expression for the
shape of the singularity:
\begin{eqnarray}
F_1(x) &=& 1-\frac{2}{3} {\rm Re} \left[ 1- {ix}
\left(
\sqrt[3]{1+y(x)}+
\sqrt[3]{1-y(x)} 
\right) 
\right]^{-1} \nonumber\\ y(x) &=& \sqrt{1+i x^3 }.
\label{1d-dos}
\end{eqnarray}
Functions $\sqrt[3]{z}$ in Eq.~ (\ref{1d-dos}) are defined to map
complex plane $-\pi < arg(z) < \pi$ to $-\frac{\pi}{3} < arg(z) <
\frac{\pi}{3}$.

For the two-dimensional case, we obtain from Eqs.~(\ref{Fd}) and (\ref{Md}):
\begin{equation}
F_2(x) = {\rm Re} \frac{1-z(x)}
{1+z(x)}, \label{2d-dos}
\end{equation}
where $z(x)$ is the solution of the transcendental equation
\begin{equation}
z(x)=\frac{1}{-i x +{\rm ln} [4 g z(x)]}.
\label{eq:z}
\end{equation}
Shape of the singularity in this case depends on the conductance $g$
and, thus, is not universal.
However, this dependence is rather weak.
For $|\omega \pm E^*| \gg W_d $, Eqs. (\ref{1d-dos}) and
(\ref{2d-dos}) match the perturbative results, Eqs.~(\ref{1st-1d-dos})
and (\ref{2nd-2d-dos}). Found energy dependence of the density of states
is shown in Fig.~\ref{Fig:dos}.
{\narrowtext
\begin{figure}[ht]
\vspace{0.4 cm}
\epsfxsize=7.7cm
\centerline{\epsfbox{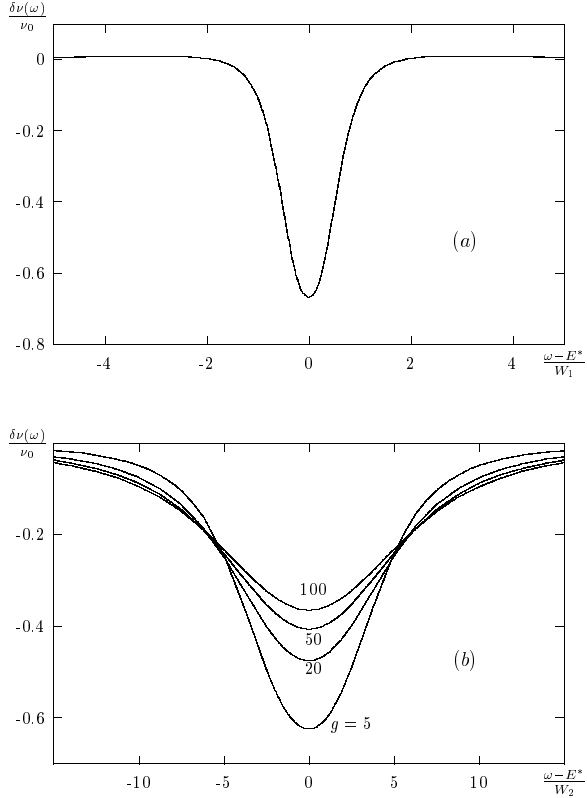}}
\vspace{0.4 cm}
\caption{Singularity in DoS for spin-down polarized electrons for (a)
$1D$, and (b) $2D$ systems. Widths of the singularity $W_{1,2}$ are
given by Eq.~(\protect\ref{Wd}), and the shape is defined by
Eqs.~(\protect\ref{DoSfinal}), (\protect\ref{1d-dos})
and (\protect\ref{2d-dos}).
}
\label{Fig:dos}
\end{figure}
}
\widetext

\section{Qualitative discussion}
\label{sec:qd}

In this section we will present a simple qualitative interpretation of
the main results obtained in the previous sections. We believe that
this simplified way of thinking provides instructive physical
intuition even though it fails to give completely quantitative
description.

It has been already emphasized in the beginning of section
Sec.~\ref{sec:0d} that the ground state of the system above the
paramagnetic limit has the structure similar to that of a
non-interacting system. All the mixing of the noninteracting states
caused by the interaction part of the Hamiltonian (\ref{ham}) is
perturbative. We can neglect it completely in a rough approximation
and consider the electrons occupying orbitals with orbital energies
$\epsilon_i < -E_Z/2$, see Fig.~\ref{Fig:2.1}, to be ``frozen''. As
soon as the spin down electron tunnels onto an orbital $0< \epsilon_0
<E_Z/2$, see Fig.~\ref{Fig:2.1}, the electron pair on this orbital is
created. Due to the interaction, this pair can mix with all the empty
states, $\epsilon_i >E_Z/2$. It is this mixing which gives rise to the
singularity in the DoS. On the other hand, within the same
approximation, all the electrons on the orbitals $\epsilon_i < -
E_Z/2$ can still be considered as ``frozen''.  ( This approximation is
similar in spirit to the well-known Cooper procedure\cite{Cooper57}.)

Thus, we arrive to the following recipe for the evaluation of the
energy of  one electron excitation. First we have to find the
eigenenergies $E^j_{(2)} (\epsilon_0)$ of  the two-electron problem
within the Hilbert space consisting of orbital $\epsilon_0$ and of all
the orbitals $\epsilon_i > E_Z/2$, see Fig.~\ref{Fig:4.1}.
(This energy spectrum naturally depends on
$\epsilon_0$ as parameter.) 
Then, the energies  $E^j_\downarrow
(\epsilon_0)$ of the one particle excitation, corresponding to
the introduction of electron onto the orbital $\epsilon_0$ are
\begin{equation}
E^j_\downarrow (\epsilon_0) =
E^j_{(2)} (\epsilon_0) - \left(\epsilon_0 - \frac{E_Z}{2}\right), 
\label{eq:4.1}
\end{equation}
since the total energy of the electron which occupied this orbital
before the tunneling event was $\epsilon_0 - E_Z/2$, while the state of the
rest of the electrons did not change.  
Accordingly, the density of states for spin-down electrons is given by
\begin{equation}
\nu_\downarrow (\omega ) \simeq \sum_{j,\epsilon_0}
\delta \left[\omega - E^j_\downarrow
(\epsilon_0)\right].
\label{eq:4.2}
\end{equation}

\narrowtext
{\begin{figure}[ht]
\vspace{0.5cm}
\epsfxsize=5cm
\hspace*{0.5cm}
\epsfbox{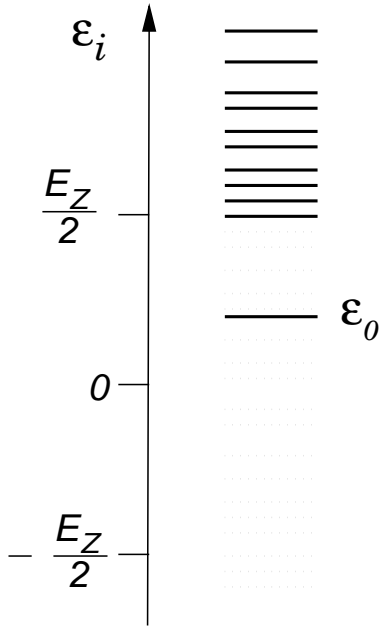}
\vspace{0.9 cm}
\caption{ Hilbert space for the solution of two - electron
problem. All the orbitals (dotted lines) with $\epsilon_i < -E_Z/2$
are excluded since they are occupied by ``frozen'' electron pairs and
the orbitals $- E_Z/2 < \epsilon_i < E_Z/2, \epsilon_i \neq
\epsilon_0$ are excluded because single occupied orbitals are not
affected by the interaction (\protect\ref{ham}).  }
\label{Fig:4.1}
\end{figure} 
}
\widetext

Now, we have to find the spectrum of two electron eigenenergies
$E^j_{(2)} (\epsilon_0)$. Since the interaction in the Hamiltonian
involves only the spin-singlet orbitals, the wave function of the
electron pair $\psi$ can be labeled by one orbital index and it is
governed by the Schr\"{o}dinger equation:
\begin{equation}
E^j_{(2)}\psi_i = 2 \epsilon_i \psi_i -\lambda \bar{\delta} \sum_{j}
\psi_{j}.
\end{equation}
The eigenenergies $E^j_{(2)}$ can be determined from the following
equation
\begin{equation}
\frac{\bar{\delta}}{2\epsilon_0 - E^j_{(2)}} 
+ \sum_{\epsilon_i > E_Z/2}
\frac{\bar{\delta}}{2 \epsilon_i - E^j_{(2)}} = \frac{1}{\lambda}.
\label{eigen}
\end{equation}
For low-lying eigenstates $E^j_{(2)} < E_Z$, one can substitute 
the summation in Eq.~(\ref{eigen}) by the integration.
Given the high-energy cut-off $\omega_c$, it yields
\begin{equation}
\frac{2 \bar{\delta}}{2\epsilon_0 - E^j_{(2)}}
=\ln\left(1+
\frac{E_Z- E^j_{(2)}-\Delta_b}{\Delta_b} \right),
\label{eigen1}
\end{equation}
where $\Delta_b = \omega_c \exp{(-2/\lambda)}$ is the binding energy
of the Cooper pair.

As we will see in a moment, $\left|2\epsilon_0 - E^j_{(2)}\right|\gg
\bar{\delta}$, and, therefore, the logarithm in Eq.~(\ref{eigen1})
should be also small and it can be expanded in the Taylor series, $\ln
(1+x) \approx x,\ |x| \ll 1$. Equation ~(\ref{eigen1}) is 
immediately simplified to
\[
\frac{2 \bar{\delta}}{2\epsilon_0 - E^j_{(2)}}
=
\frac{E_Z - \Delta_b - E^j_{(2)}}{\Delta_b},
\]
and we obtain the solution for two relevant eigenenergies: 
\begin{equation}
 E^\pm_{(2)}(\epsilon_0)
 = \epsilon_0 +\frac{\Omega_b}{2} \pm 
\sqrt{\left(\frac{\Omega_b}{2}-\epsilon_0\right)^2
+ 2 \bar{\delta} \Delta_b }.
\label{eq:Ej}
\end{equation}
All the other two-electron states have  energies larger than $E_Z$ and
they are not important for us.
In Eq.~(\ref{eq:Ej}) energy 
\begin{equation}
\Omega_b = E_Z-\Delta_b
\label{omegab}
\end{equation}
has the meaning of the energy of the bound state of the Cooper pair
measured from the Fermi level. It plays the role of energy $\Omega$
from Eq.~(\ref{omega}) in the rigorous solution. We will return to the
discussion of the discrepancy between Eq.~(\ref{omegab}) and
Eq.~(\ref{omega}) later.

Substituting Eq.~(\ref{eq:Ej}) into Eq.~(\ref{eq:4.1}), we obtain
the energy of one particle excitations
\begin{equation}
E_{\downarrow}^\pm(\epsilon_0) = E_b^* \pm \sqrt{
\left(\frac{\Omega_b}{2}-\epsilon_0\right)^2 + 2 \bar{\delta} \Delta_b
},
\label{down}
\end{equation}
where the position of the singularity
\begin{equation}
E_b^* = \frac{E_Z + \Omega_b}{2}
\label{naive}
\end{equation}
is similar to the energy $E^*$ in exact Eq.~(\ref{singular}) up to the
substitution $\Omega \to \Omega_b$.

According to Eqs.~(\ref{eq:4.2}) and Eq.~(\ref{down}), the density of
states {\em vanishes} in the energy strip $|E-{E}^\ast_b|<
\left(2\bar{\delta}\Delta_b\right)^{1/2}$ -- hard gap in the DoS is
formed, compare with the exact result, Eq.~(\ref{0d-dos}). 
The origin of this tunneling anomaly is the avoided
crossing of the two-electron state formed by the tunneling electron
(energy $2\epsilon_0$) with the bound state of the Cooper pair (energy
$\Omega_b$).

It is also noteworthy that if a spin-up electron tunnels into the grain,
it never finds the pair for itself, and, therefore, no tunneling anomaly
happens in this case. It means that the overall DoS
does not vanish but rather shows the suppression by a factor of
two. However, for the spin-polarized electrons
tunneling into the grain, we predict the complete suppression of the
tunneling DoS.   

The same arguments allow to justify the similar singularity, when
spin-up electron with energy $-E_Z/2 < E < 0$ tunnels out from the
system, while the spin down electrons tunneling from the system are
not affected.

The qualitative consideration above grasps the correct physics,
however it fails to describe the effect quantitatively, it predicts
correctly neither the position  nor the width of
the gap.  This is similar to the discrepancy between the binding energy
$\Delta_b$ in the original Cooper procedure and the correct BCS gap $\Delta$:
all the electrons below the Fermi energy were frozen. To remedy this
drawback, one has to employ a parametrically exact procedure
described in Secs. \ref{sec:0d} and \ref{sec:hd}.

Let us now discuss the results obtained for the disordered bulk
systems, Sec.~\ref{sec:hd}. They can be briefly summarized as follows:
(i) Singularity in the DoS persists; (ii) its position does not
change; (iii) energy scale of the singularity depends on both disorder
and dimensionality, see Eq.~(\ref{Wd}).

In order to understand the physics behind the singularities in the
bulk systems, let us recall the meaning of zero-dimensional
approximation. Strictly speaking, it implies that during the time $t_E
\simeq \hbar/E$, (where $E$ is the energy scale relevant for the
problem), diffusively moving electron can visit all the system. The
characteristic time of this diffusion is $\simeq L^2/D$, ($L$ is the
characteristic size of the system) which means
that zero-dimensional approximation is applicable provided that the
energy scale $E$ does not exceed the Thouless energy $E_c=\hbar D/L^2$.
In our problem the relevant energy scale is the gap width $W_0$ from
Eq.~(\ref{W_0}), and condition
\begin{equation}
1> \frac{W_0}{E_c} \propto L^{2 - d/2} 
\label{eq:4.3}
\end{equation}
is definitely violated for the infinite systems $L \to \infty$, (here
$d=1,2$ is the dimensionality of the system).

It is clear, that as soon as the condition (\ref{eq:4.3}) breaks down
the geometrical size of the system $L$ as well as its mean level
spacing $\bar{\delta}= 1/(\nu_0 L^d)$ is not relevant since the electron
can not diffuse during finite time $t_E = \hbar/E$ over the distance
larger than $L_E = \sqrt{Dt_E}$. On the other hand, it effectively
visits all the space on the scale smaller than $L_E$. Therefore, the
following approximation holds: in order to understand the properties
of the diffusive system associated with the energy scale $E$, we can
separate the system into smaller patches of the size $L_E =
\sqrt{\hbar D/E}$, and, then, apply zero-dimensional description to
each patch independently, (assuming that different patches do not
``talk'' to each other).

Let us now apply this strategy to the problem in hand. First, we
notice that the position of the singularity in $0D$ grain
(\ref{singular}) does not depend on the size of the grain and
therefore the singularity in each patch should be at the same energy
$E^*$ as in zero-dimensional systems. Second, level spacing
$\bar{\delta}$ entering into the width of the singularity
Eq.~(\ref{W_0}) does depend on the size of the patch
\begin{equation}
\bar{\delta} = \frac{1}{\nu_0 L_E^d}.
\label{eq:4.4}
\end{equation} 
In this formula, scale $L_E$ is itself determined by the width of the
singularity, $E \simeq W_d$, so that the scale $W_d$ has to be
determined self-consistently. Substituting Eq.~(\ref{eq:4.4}) into
Eq.~(\ref{W_0}), we find
\begin{equation}
W_d \simeq \left(\frac{\Delta^2}{\nu_0 L_W^d \Omega}\right)^{1/2}; 
\quad L_W \simeq \left(\frac{D}{W_d}\right)^{1/2}.
\label{eq:4.5}
\end{equation}
Solving Eq.~(\ref{eq:4.5}), we obtain
\begin{equation}
W_d \simeq \Delta \left(\frac{\Delta^{d/2}}{\nu_0 \Omega
D^{d/2}}\right)^{\frac{2}{4-d}},
\label{eq:4.6}
\end{equation}
which agrees with the rigorous results (\ref{Wd}) for one- ($d=1$) and
two- ($d=2$) dimensional systems. However, the quantitative
description requires machinery like one used in Sec.~\ref{sec:hd}.

\section{Relevant perturbations}
\label{sec:rp}
\subsection{Spin-orbit scattering}

In our previous consideration we assumed that electronic spin is
a good quantum number, i.e., impurity  scattering of electrons does not
cause spin-flips.
There are two sources of spin relaxation of conduction electrons:
localized spins (paramagnetic impurities) and spin-orbit (SO) scattering
of electrons by non-magnetic disorder. The latter is characterized by the
scattering amplitude 
\begin{equation}
iv_{so}([\mbox{\boldmath $p$}_f 
\times \mbox{\boldmath $p$}_i]\cdot  \mbox{\boldmath $\sigma$})/p_F^2,
\label{amp}
\end{equation}
where $\mbox{\boldmath $p$}_f$ and $\mbox{\boldmath $p$}_i$ are final
and initial momenta of an electron, and $\mbox{\boldmath $\sigma$}$ is
the spin operator $\mbox{\boldmath $\sigma$} =
\left(\hat{\sigma}_x,\hat{\sigma}_y,\hat{\sigma}_z\right)
$ 
whose components are
Pauli matrices. It acts on the spinor wave function of the electron.

Let us discuss the effect of SO scattering first starting with the
qualitative physical picture.
In the absence of both SO interactions and magnetic field two spin
states which belong to a given orbital have the same energy.
Magnetic fields splits this degeneracy.
It was important for us that the splitting $E_Z$ is exactly the same for
all of the orbital states.
Now let us turn on SO interaction.
Without external magnetic field the states are still double degenerate
due to the T-invariance (Kramers doublets\cite{Kramers}).
Magnetic field is well known to split the Kramers doublets similar to
how it splits pure spin states in the absence of SO interactions. 
The main difference is that this splitting is not exactly uniform any
more (see e.g. Ref.~\onlinecite{Kravtsov}).
It is this dispersion of splittings that smears the DoS singularity.
Zeeman splitting dominates the magnetic field effect on
superconductivity only provided SO interaction is weak. 
However the DoS singularity turns out to be sensitive even to a weak SO
scattering, since the characteristic SO energy(dispersion is splitting
of Kramers doublets) should be compared with the width of the
singularity $W_d$ rather than with the splitting $E_Z$ itself.
The DoS for finite SO scattering can be evaluated in a way similar to
our calculation in Sec.~\ref{sec:hdn}.

Cooperon (or vertex) is formed by two electron Green functions. In the
absence of external magnetic field it is convenient to classify Cooper
poles by the total spin of two electrons $ \mbox{\boldmath $S$}_+=
(\mbox{\boldmath $\sigma$}_1+ \mbox{\boldmath $\sigma$}_2)/2 $.
Spin-orbit scattering does not affect the spin singlet part of the
Cooperon ($\mbox{\boldmath $S$}^2_+ = 0$) because the spin-orbit
scattering preserves $T$- invariance.  However, this scattering leads
to total spin relaxation, i.e., triplet ($\mbox{\boldmath $S$}^2_+ =
2$) component of Cooperon decays (pole in $\omega$-plane is shifted
from the real axis even for $Q=0$)\cite{LHN}.

External magnetic field is coupled with the difference 
$
\mbox{\boldmath $S$}_-=
(\mbox{\boldmath $\sigma$}_1- \mbox{\boldmath $\sigma$}_2)/2
$ of two electron spins, and as a
result we classified Cooperon by the eigenvalue of the operator 
$
\mbox{\boldmath $S$}_-
\cdot
\mbox{\boldmath $E$}_Z
$.  
These eigenvalues for $\mbox{\boldmath $S$}_-^2=2$ are $E_Z,\ 0,\
-E_Z$ corresponding to $S_-^z =1,\ 0,\ -1$ and $0$ for 
 $\mbox{\boldmath $S$}_-^2=0$. Neither
of those two classifications is exact when both magnetic field and SO
scattering take place simultaneously: operators $S^2_+$ and $S^z_-$ do
not commute.  On the other hand, as it was already mentioned, we
should assume that SO effect is weak.  This allows us to evaluate
Cooperon perturbatively.

Summing usual ladder diagrams  and taking additional scattering
amplitude Eq.~(\ref{amp}) into account, we end up with an equation 
for $4\times 4$ matrix of the Cooperon
which we already discussed qualitatively:
\begin{equation}
\left[\left(-i\omega+DQ^2\right)\hat{I} +
i \mbox{\boldmath $E$}_Z\cdot \mbox{\boldmath $S$}_-
+
\frac {2\mbox{\boldmath $S$}_+^2}{3\tau_{so}}\right]
{\hat C} =
\frac{\hat{I}}{\tau},
\label{soeq}
\end{equation}
where $\tau_{so}=1/(2\pi \nu_0 v_{so}^2)$ is the time of the spin
relaxation by SO scattering, matrix $\hat{I}$ is the direct product
$\hat{I} =\sigma_0^s \otimes \sigma_0^e$, where unit matrix
$\sigma_0^s$ is acting in the spin $2\times 2$ space, and $\sigma_0^e$
is a unit matrix in $2\times 2$ space of the electron lines.
Operators $\mbox{\boldmath $S$}_\pm$ are defined as $2\mbox{\boldmath
$S$}_+=\left( \mbox{\boldmath $\sigma$}^s + \mbox{\boldmath $n$}
\sigma_0^s \right)\otimes \sigma_0^e$ and $2\mbox{\boldmath
$S$}_-=\left(\mbox{\boldmath $\sigma$}^s - \mbox{\boldmath $n$}
\sigma_0^s\right) \otimes \sigma_z^e$, where $ \mbox{\boldmath $n$}=
\left(1_x,1_y,1_z\right)$ is the unit vector, and $ \mbox{\boldmath
$\sigma$}^s= \left(\sigma_x^s,\sigma_y^s,\sigma_z^s\right)$ are the
Pauli matrices in the spin space.

Instead of diagonalizing Eq.~(\ref{soeq}), we can just evaluate correction to
$C^{-1}_{\uparrow (\downarrow)}$,
\begin{equation}
C^{-1}_{\uparrow} \equiv
\left[C^{-1}\right]^{\uparrow,\downarrow}_{\uparrow, \downarrow} \quad
C^{-1}_{\downarrow}\equiv \left[C^{-1}\right]_{\downarrow,
\uparrow}^{\downarrow,\uparrow}
\end{equation}
in the  first order of the perturbation theory in $(E_Z \tau_{so})^{-1}
\ll 1$. 
This correction turns out to be $2/(3 \tau_{so})$.
As a result, SO scattering simply shifts the Cooper poles
$C_{\uparrow(\downarrow)}$:
\begin{equation}
C_{\uparrow(\downarrow)}(\omega,Q) = \frac{1}{-i\omega \mp i E_Z +DQ^2 
+ \frac{2}{3\tau_{so}} }.
\label{CSO}
\end{equation}

From Eq.~(\ref{CSO}) one can guess that the all the interesting
results can be obtained from the results of Secs.~\ref{sec:0d} and
\ref{sec:hd} by substituting 
\begin{equation}
\omega \to \omega +i\Gamma,
\label{rule}
\end{equation}
where $\Gamma$ is the spin orbit rate
\begin{equation}
\Gamma=\frac{2}{3\tau_{so}},
\label{etaso}
\end{equation}
so that the final result for the density of states is
\begin{equation}
\frac{\nu_{\uparrow(\downarrow)}(\omega )}{\nu_0}
={\rm Re} F_d \left( \frac{\omega\pm E^* + i\Gamma}{W_d} \right)
\label{mfd}
\end{equation}
with $W_d$ and $E^*$ determined by Eqs.~(\ref{W_0}),
(\ref{Wd}), and (\ref{E^*}),
and dimensionless function $F_d$ are given by Eqs.~(\ref{0d-dos}),
(\ref{1d-dos}) and (\ref{2d-dos}).
We notice that the DoS singularity gets smeared by SO
scattering when $\Gamma \gtrsim W_d$. This is in contrast with conventional
superconductivity which is known to be stable with respect to SO
scattering since the latter does not violate T-invariance.

This guess turns to be correct. In order to demonstrate it one has to show
that not only the Cooperon but also Hikami boxes from
Sec.~\ref{sec:hdn} are modified according the rule
(\ref{rule}).
Taking into account the spin orbit scattering Eq.~(\ref{amp})
in impurity lines on diagrams Figs.~\ref{Fig:3.6} and \ref{Fig:3.7}, 
we obtain
\begin{eqnarray*}
B^{(2)}(\omega_1,\omega_2,Q_1,Q_2)
 = \frac{1}{(2\pi \nu_0)^3}\left[-i(\omega_1+\omega_2)+
D(Q_1^2+Q_2^2)+\frac{4}{3\tau_{so}} \right], \\
B^{(3)}(\omega_1,\omega_2,\omega_3;Q_1,Q_2,Q_3) = 
-\frac{2}{(2\pi \nu_0)^5} \left[-i(\omega_1+\omega_2+\omega_3)
+D(Q_1^2+Q_2^2+Q_3^2) +2\tau_{so}\right].  
\end{eqnarray*}
With the same rigor as in Sec.~\ref{sec:hdn}, we conjecture that
\[
B^{(n)} \{ \omega_j, Q_j \}= \frac{C_n}{(2\pi\nu_0)^{2n-1}} 
\sum_{j=1}^n \left(-i\omega_j+DQ_j^2+\frac{2}{3\tau_{so}} \right),
\]
therefore, rule (\ref{rule}) is satisfied which gives Eq.~(\ref{mfd}).

Finally we emphasize that, in Eq. (\ref{mfd}) complex argument of the
function $F_d(z)$ should be located on the physical sheet:
$z=|z|\exp{i\varphi}$ and $-\pi < \varphi < \pi$.  Thus, the functions
$\sqrt[d]{z}$ in Eqs. (\ref{0d-dos}) and (\ref{1d-dos}) are defined to
map complex plane $-\pi < arg(z) < \pi$ to $-\frac{\pi}{d} < arg(z) <
\frac{\pi}{d}$. 

\subsection{Paramagnetic impurities, orbital effects of the magnetic
field}

The derivation in the previous section suggests that
any physical mechanisms of violation of either T-invariance or
conservation of spin will have similar effect on the DoS singularity.
Indeed, in 0D and 1D cases in the presence of magnetic field $H$ and
paramagnetic spins, Eq. (\ref{cooper}) takes the following form\cite{AA}
\begin{equation}
C(\omega,Q)=\frac{1}{-i\omega+DQ^2+\frac{1}{\tau_{tot}}},
\label{coopertot}
\end{equation}
where $\tau_{tot}$ is a combination of phase and spin relaxation effects
\begin{equation}
\frac{1}{\tau_{tot}}=\frac{1}{t_{\varphi}} + \frac{1}{t_s}.
\label{taut}
\end{equation}
Both SO scattering and spin exchange with paramagnetic impurities
($\tau_s$) lead to spin relaxation:
\begin{equation}
\frac{1}{t_s}=\frac{2}{3} \left( \frac{1}{\tau_{so}} + \frac{1}{\tau_s}
\right).
\end{equation}
Phase may relax either due to inelastic processes or due to magnetic
field effect on orbital motion of electrons:
\begin{equation}
\frac{1}{t_{\varphi}}= \frac{1}{\tau_{\varphi}} + \frac{1}{\tau_H}.
\label{tauphi}
\end{equation}

When the transverse dimension of the wire (size of the grain) $a$,
exceeds man free path $l$, $\tau_H$ can be estimated as\cite{AA,LR}
\begin{equation}
\frac{1}{\tau_H}=A\frac{\Omega_H^2}{E_T} \propto
D\left(\frac{aH}{\phi_0}\right)^2.
\label{tauh}
\end{equation}
Here $\phi_0=\frac{hc}{2e} \simeq 2\times 10^{-7} Gs\cdot cm^2$ is the
superconductivity magnetic flux, $E_T$ is the "transverse" Thouless
energy, $E_T=D/a^2$, and $\Omega_H$ is the Cooperon "cyclotron
frequency" (cyclotron frequency for a particle which mass and charge
equal to $(2D)^{-1}$ and $2e$ respectively)\cite{AA}.
\begin{equation}
\Omega_H=\frac{4DeH}{\hbar c}
\label{Omegah}
\end{equation}
Numerical coefficient $A$ is not universal: it depends on the geometry
of the superconductor, on the direction of magnetic field, etc.

Equation (\ref{coopertot})--(\ref{Omegah}) are valid also for 2D films
provided $H$ is parallel to the film plane.
In this case
\begin{equation}
\frac{1}{\tau_H} =\frac{a^2 \Omega_H^2}{48D}=
\frac{D(eHa)^2}{3c^2 \hbar^2}.
\end{equation}
Now, we again conjecture that Eq. (\ref{hikamin}) for Hikami boxes is
still valid with addition of $1/\tau_{tot}$ to all $DQ_j^2$.
If so, Eq. ({\ref{mfd}) for DoS is still valid, but instead of 
Eq.~(\ref{etaso}) we should substitute
\begin{equation}
\Gamma = \frac{1}{\tau_{tot}}
\label{etatot}
\end{equation}
Rate $\tau_{\varphi}^{-1}$ in Eq. (\ref{tauphi}) 
is the contribution of inelastic collisions. 
This contribution will be estimated in the next section.
Evolution of the density of states with the rate $\alpha= \Gamma/W_d$ for
different dimensions is shown in Fig.~\ref{Fig:5.1}.
{\narrowtext
\begin{figure}[ht]
\epsfxsize=6cm
\centerline{\epsfbox{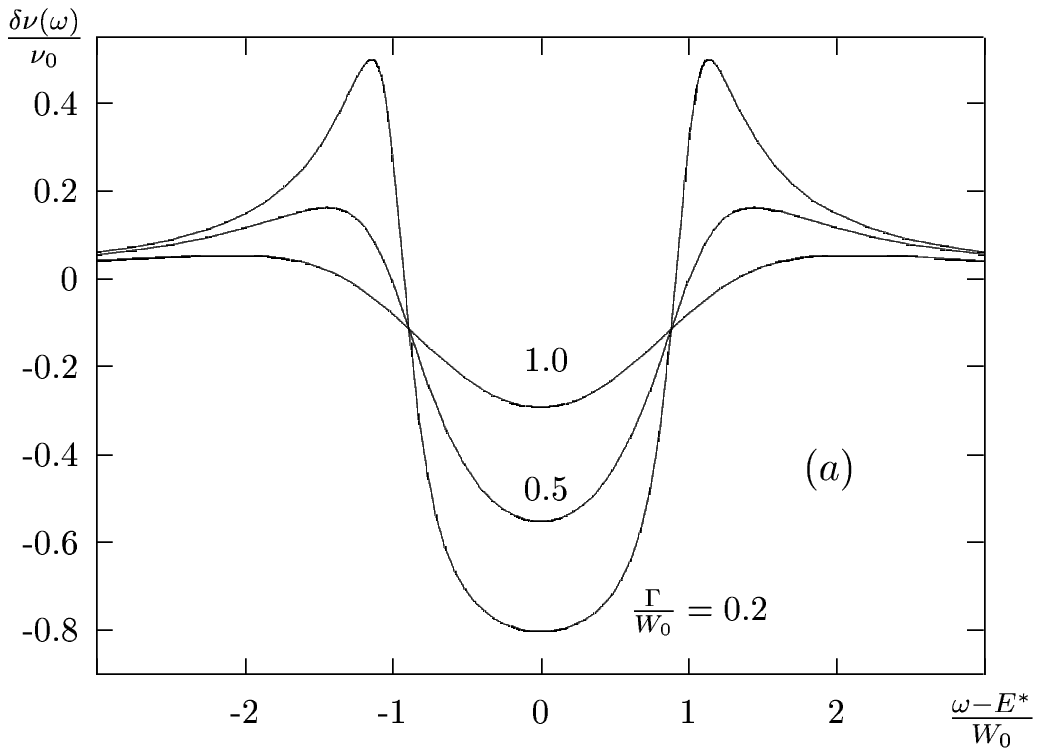}}
\epsfxsize=6cm
\centerline{\epsfbox{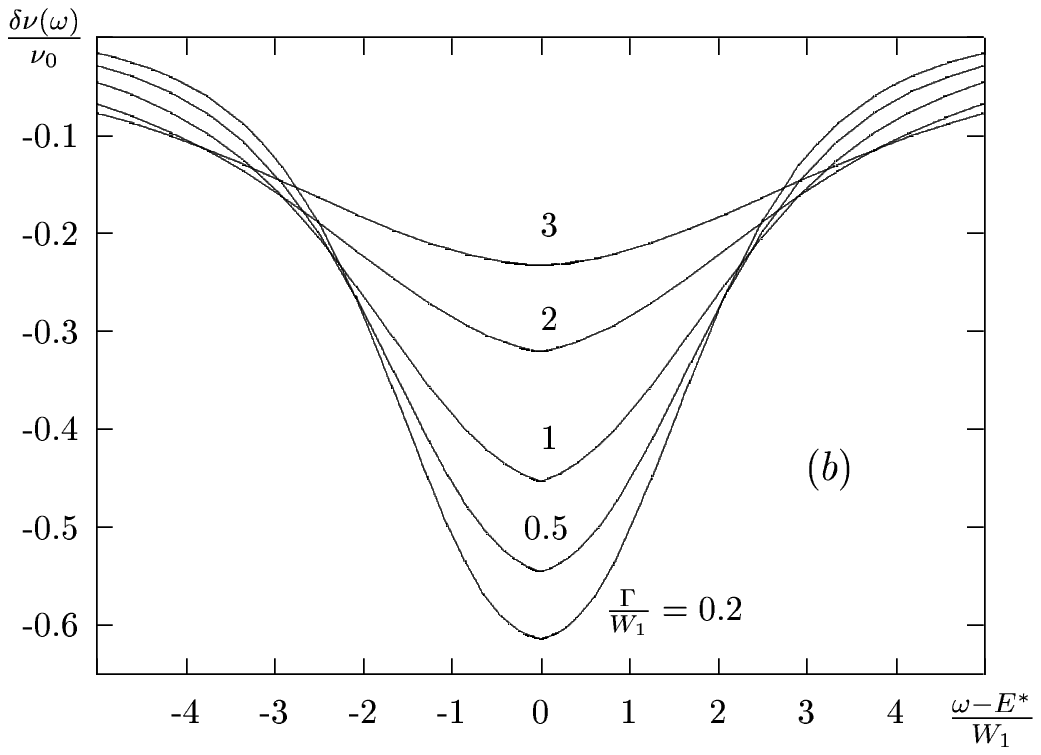}}
\epsfxsize=6cm
\centerline{\epsfbox{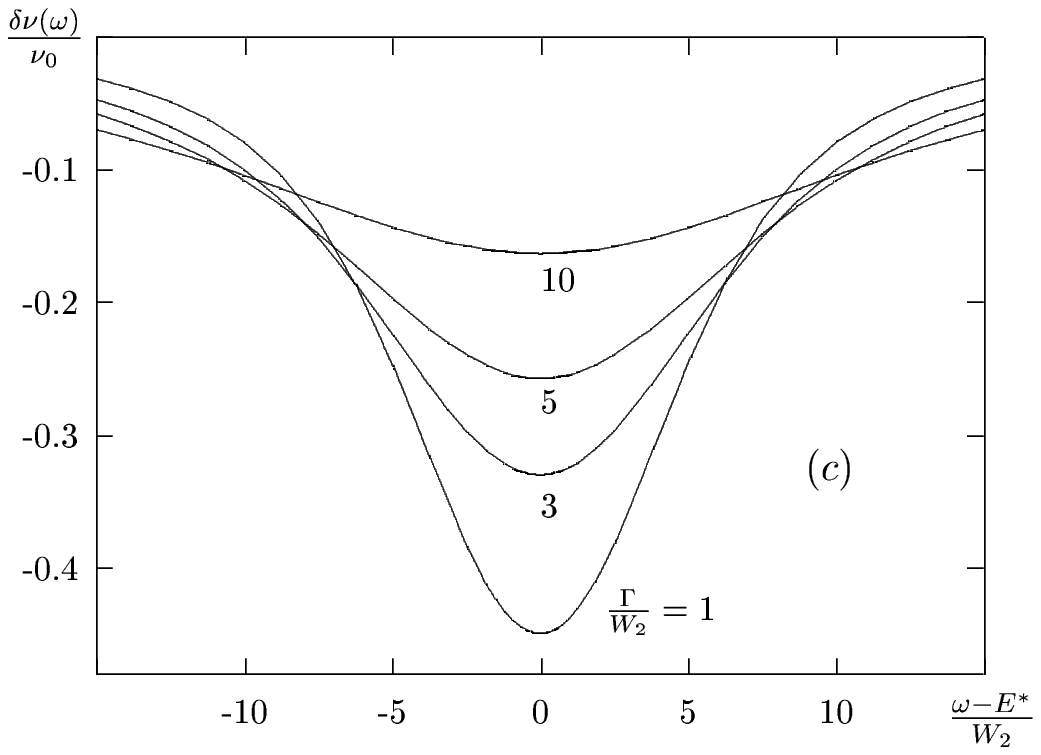}}
\vspace{0.25 cm}
\caption{Singularity in DoS for spin-down polarized electrons for (a)
$0D$, and (b) $1D$ and (c) $2D$ systems for different values of
dimensionless rate $\alpha = \Gamma/W_d $. The DoS in two dimensional
case is plotted for conductance $g=10$. 
}
\label{Fig:5.1}
\end{figure}
}
\widetext

\subsection{Finite temperature and inelastic processes}

Our previous consideration, strictly speaking, applies only when
temperature $T$ equals to zero. Let us now discuss effects of
finite $T$.

Temperature manifests itself through distribution of electrons in
energy. This distribution substantially depends on $T$ only for
energies of the order of $T$. On the other hand, for tunneling
anomalies at finite bias $eV \sim E^*$, Eq. (\ref{E^*}), only low
temperature region $T \ll E^*$ (recall that the Zeeman splitting $E_Z$
as well as superconducting gap $\Delta$ are normally of the order of
$E^*$) is of interest.  Indeed, thermal broadening of the Fermi steps
in the leads washes out any singularity as soon as $T$ exceeds the
width $W$ of this singularity.  Since the widths $W_d$, Eqs.~
(\ref{W_0}) and (\ref{Wd}), are much smaller than $E^*$, there will be
already no anomaly of the tunneling current at $T \sim E^*$.  On the
other hand, at $T \ll E^*$ the equilibrium occupation number of a
state with the energy of the order of $E^*$ is exponentially small and
thus can be neglected.  Therefore, formally determined DoS,
Eq. (\ref{DoSfinal}) is temperature independent up to terms of the
order of $\exp(-E^*/T)$.  However, the observable, namely tunneling
conductance $\sigma_T(eV)$, depends on $T$ through the Fermi
distribution $n_F(\omega/T)$ in the leads. To evaluate the singular
part $\delta \sigma_T(eV)$ of the tunneling conductance, one has to
convolute $\delta \nu (\omega)$ with the derivative of the biased
Fermi distribution:
\begin{equation}
\frac{\delta \sigma_T(eV)}{\sigma_T^0} = -\int d\omega \frac{\partial
n_F(\frac{\omega-eV}{T})}{\partial \omega} \frac{\delta
\nu_{\uparrow}(\omega)+\delta \nu_{\downarrow} (\omega) }{\nu_0}
\label{cond}
\end{equation}

Substitution of Eq.~(\ref{DoSfinal}) into Eq. (\ref{cond}) yields
for $\sigma_T= \sigma_T^0 + \delta \sigma_T$:
\begin{equation}
\frac{\sigma_T(eV)}{\sigma_T^0} = \int \frac{d\omega}{4 T
\cosh^2{(\frac{\omega-eV}{2T})}} \sum_{\sigma=\pm 1} {\rm Re} F_d \left(
\frac{\omega+\sigma E^*+i\Gamma}{W_d} \right)
\label{cond2}
\end{equation}

To complete this discussion, let us estimate contribution of inelastic
collisions of electrons, $1/\tau_{\varphi}$ Eq.~(\ref{tauphi}), to
$\Gamma$.  Since we are dealing with rather highly excited states
($\omega \sim E^* \gg T$), relaxation rate $1/\tau_{\varphi}$ is
determined by large energy transfer ($\sim E^*$), and thus is
temperature independent.  In all interesting cases $1/\tau_{\varphi}$
simply coincides with the energy relaxation rate $1/\tau_{\epsilon}$.
The latter (in metals, for reasonably low energies $\omega$) is
determined by inelastic collision between electrons (see
Ref.~\onlinecite{AA} for review) and can be estimated as
\begin{equation}
\frac{1}{\tau_{\varphi}} \simeq \frac{1}{\tau_{\epsilon}}
\sim \frac{1}{\nu_0 L^d_{\omega}} = \frac{\omega}{g(L_{\omega})},
\end{equation}
where $L_{\omega}=\sqrt{D/\omega}$ is the length of diffusion in time
$\omega^{-1}$ and $g(L)= \nu_0 DL^{d-2}$ is conductance of the
$d$-dimensional sample with size $L$. For zero dimensional case the
inelastic process of this type can be neglected completely since 
$\frac{1}{\tau_{\varphi}}$ compares with the level spacing only at the
energies $E^\ast$ of the order of the Thouless
energy\cite{Zinoviev,Aronov}.

Using Eqs.~(\ref{eq:4.6}) we can estimate dimensionless product $W_d
\tau_{\varphi}$ as
\begin{eqnarray}
W_d \tau_{\varphi} &\sim& \left( \frac{\Delta}{\Omega}
\right)^{\frac{2}{4-d}} g(\xi)^{\frac{2-d}{4-d}} \nonumber\\
&\sim& \left( \frac{\Delta}{\Omega} \right)^{\frac{2}{4-d}}
\times \left\{ \begin{array}{ll} g^{1/3}(\xi) & \mbox{d=1}\\ 
                                \log{g(\xi)} & \mbox{d=2},
               \end{array}
       \right.
\end{eqnarray}
where $\xi=L_{\Delta}$ is the coherence length.
Since $\Delta \sim \Omega$, this estimation implies that as long as
localization length $L_{loc}$ exceeds $\xi$ the width $W_d$ is much
bigger than $1/\tau_{\varphi}$ and inelastic collisions are
irrelevant($L_{loc}$ can be estimated from the equation $g(L_{loc}) \sim
1$).

Together with Eqs.~ (\ref{W_0}) and (\ref{Wd}) 
 for $W_d$ and Eqs.~(\ref{taut})--(\ref{etatot})
for $\Gamma$, Eq.~(\ref{cond2}) completely describes tunneling anomalies
at $eV \sim \pm E^*$ for $W_d \geq T, \Gamma$ and $d=0,1$.
For $d=2$, Eq.~(\ref{cond2}) is valid only provided that the external
magnetic field is parallel to the plane. 
The orbital effect of this perpendicular component on the tunneling
anomaly is discussed in the next section. 

\subsection{Magnetic field perpendicular to the film}


It was already mentioned that for zero- and one-dimensional cases
orbital effects of magnetic field manifest themselves through addition
$1/\tau_H$ from Eq.~(\ref{tauh}), to the parameter $\Gamma$, see Eqs.~
(\ref{etaso}), (\ref{Omegah}), and (\ref{etatot}).  The same is true
in two dimensions provided the magnetic field is parallel to the film
plane\cite{AA}.  However, effect of a component of magnetic
field perpendicular to the plane requires a separate
consideration.

Similarly to usual calculation of anomalous
magnetoresistance\cite{LR,AA}, we need to derive and to solve equation
for the Cooperon in perpendicular magnetic field rather than to take
this field into account perturbatively.  This equation is well known
to be a Schr\"{o}dinger equation in imaginary time for a particle with
charge $2e$ and mass $1/(2D)$ in the magnetic field $H$.  For
$H=0$ eigenfunctions of this equations are plane waves.  When this
field $H$ is finite but parallel to the plane and weak enough
(magnetic length $l_H = (\hbar c/eH)^{1/2}$ exceeds the film thickness
$a$, or $\Omega_H$, Eq.~(\ref{Omegah}), is smaller than
"perpendicular" Thouless energy $E_T=D/a^2$), it can be taken into
account perturbatively. Eigenfunctions remain to be plane waves, so
that Cooperon keeps its form Eq.~(\ref{coopertot}), but the
eigenvalues are shifted by $\tau_H^{-1}$.

Contrarily, even weak perpendicular component of ${H}$ changes
eigenfunctions, and as a result the form of the Cooper pole is also
modified.  Each eigenfunction should be characterized by the number of
Landau band $n$ and by one of the components of momentum $Q$ (one of
the coordinates of the guiding center) rather than by both
components of the momentum.  Corresponding eigenvalue equals to
$\Omega_H (n+1/2)$, where $\Omega_H$ is given by Eq.~(\ref{Omegah}),
{\em i.e.}, it is determined by the number of Landau band, and it is
independent of the location of the center.

As a result, in all previous calculations $DQ_j^2$ should be
substituted by $\Omega_H (n+1/2)$, and instead of integrating over
$(dQ_j)$, we have to sum over $n$ and divide the result by $4\pi
l_H^2$.  At $H=0$ integration over $dQ$ is limited from above by $DQ^2
\lesssim \Delta $, see discussion after Eq.~(\ref{Md}).  For the same
reason we sum over $n$ from 1 till $N \simeq\Delta/\Omega_H$.  In
order to evaluate $\delta \nu_{\sigma}(\omega)$, we have to perform
calculation similar to that of Sec.~\ref{sec:hd} with such changes.

Equation (\ref{corrdos}) for the DoS should be rewritten as
\begin{equation}
\frac{\delta \nu_{\sigma}(\omega)}{\nu_0} = -\frac{\Delta^2}{\nu_0 \Omega}
{\rm Im} \frac{\partial}{\partial \omega}
\int^1_0 d\eta \frac{1}{4 \pi l_H^2} \sum_{n=1}^{N}
 \gamma_{\omega} (\omega_{n\sigma}, n),
\label{nuH}
\end{equation}
where $\sigma =\pm 1$ corresponds to $\uparrow$ and $\downarrow$
respectively, and the short hand notation
\begin{equation}
\omega_{n\sigma} = 2 (\omega +\sigma E^*) + i\, \Omega_H
\left(n+\frac{1}{2}\right),
\end{equation}
is introduced.
After obvious modification of Eq.~(\ref{Eq-vertex}), we obtain instead
of Eq.~(\ref{gammaz}) 
\begin{equation}
\gamma_{\omega}(\zeta,n) =
\frac{Z(\omega)}{-i\zeta+\Omega_H (n+\frac{1}{2})+2m(\omega)},
\end{equation}
where $Z(\omega)$ and $m(\omega)$ can still be expressed through
$\beta_0$ and $\beta_1$ according to Eq. (\ref{zm}).
Consequently, functions $\beta_0(\omega)$ and $\beta_1(\omega)$ can
be connected with $\gamma_{\omega}(\zeta, n)$ by  equations similar
to Eq.~(\ref{betas}):
\begin{equation}
\beta_p(\omega) = \eta \frac{\Delta}{\nu_0 \Omega} \sum_{n=1}^{N}
\frac{\gamma^2_{\omega}(\omega_n,n)} 
{\left[ -i\omega_n+\Omega_H (n+\frac{1}{2})\right]^p}.
\label{betan}
\end{equation}
Equation (\ref{betan}) is valid for $p=0, 1$ and for the both spin
directions.  We substitute Eq.~(\ref{gammaz}) into Eqs.~(\ref{nuH})
and (\ref{betan}), carry out summation over $n$, and obtain
Eqs.~(\ref{eq:3.3.1}) and (\ref{eq:3.3.3}). The scale of the
singularity coincides with $W_2$ from Eq.~(\ref{Wd}) while
dimensionless function $M_d$ is substituted by
\begin{equation}
M_H(x) = \ln \left(\frac{\Delta}{\Omega_H}\right)
- \psi\left(\frac{1}{2} + \frac{x}{\alpha_H}\right).
\label{MH}
\end{equation}
Here, orbital effect of the magnetic field is characterized by
dimensionless parameter $\alpha_H = \Omega_H/W_2$ and
 $\psi(x)$ is the digamma function
\[
\psi(x) = \sum_{n=0}^{\infty} \left(\frac{1}{n+1}-\frac{1}{n+x}\right)
 - \mbox{\boldmath $C$},
\]
and $ \mbox{\boldmath $C$} \approx 0.577 \dots $ is the Euler constant.
If magnetic field is weak $\alpha_H \ll 1$, we can use the asymptotic
expansion $\psi (x) \approx \ln x,\ x\gg 1$ and recover
two-dimensional result $M_2$ from Eq.~(\ref{Md}). Since function $M_H$
depends on additional variable $\alpha_H$ only as on a parameter we can
use the solution of Sec.~\ref{sec:hdn} to obtain density of states:
\begin{equation}
\frac{\nu_\sigma (\omega)}{\nu_0} = F_H \left(\frac{\omega -
\sigma E^* + i\Gamma}{W_2}  \right),
\label{DoSH}
\end{equation}
where energy scale $W_2$ is given by Eq.~(\ref{Wd}),
we included previously discussed broadening mechanisms according to
rule (\ref{rule}) with the rate $\Gamma$ given by Eq.~(\ref{etatot}).
Dimensionless function $F_H(x)$ is given by
\begin{equation}
F_H(x)={\rm Re}\frac{\alpha_H - \psi^\prime \left[\frac{1}{2} + \frac{-i x + y (x)
}{\alpha_H}\right]}
{\alpha_H + \psi^\prime \left[\frac{1}{2} + \frac{-i x + y (x)
}{\alpha_H}\right]},
\end{equation}
where the function $y(x)$ is the solution of the equation
\[
y(x) =  \ln 4g - \ln \alpha_H
- \psi\left(\frac{1}{2} + \frac{-i x + y (x)}{\alpha_H}\right).
\]
The density of states in two dimensional films for different values of
the parameter $\alpha_H$ is shown on Fig.~\ref{Fig:5.2}.

{\narrowtext
\begin{figure}[ht]
\vspace{0.2 cm}
\epsfxsize=6.7cm
\centerline{\epsfbox{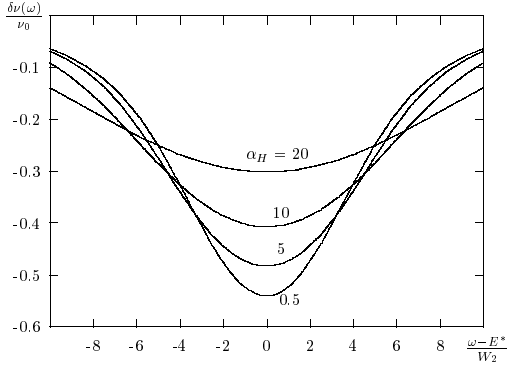}}
\vspace{0.15 cm}
\caption{Singularity in DoS of two dimensional films
for spin-down polarized electrons for  for different values of
dimensionless magnetic field $\alpha_H = \Omega_H/W_2 $. 
Curves are  plotted for conductance $g=10$. 
}
\vspace{0.15cm}
\label{Fig:5.2}
\end{figure}
}
\widetext

Closing this subsection, we present the asymptotic behavior of
$F_H(x)$ for two limiting cases. In the weak fields $\alpha_H \ll 1$
the magnetic field slightly perturbs the two-dimensional result
(\ref{2d-dos}):
\begin{equation}
F_H(x)=
F_2\left(x \right)
+\frac{1}{12}\left(\frac{\Omega_H}{W_2}\right)^2
{\rm Re} \frac{z^3(x)}{\left[1+z(x)\right]^2},
\label{as1}
\end{equation}
where dimensionless functions $F_2(x)$ and $z(x)$ are defined in 
Eqs.~(\ref{2d-dos}) and (\ref{eq:z}) respectively. 

In the opposite limit $\alpha_H \gg {\rm max}\left( 1,
\Gamma/W_2\right )$ the depth of the singularity is
controlled by solely cyclotron frequency (\ref{Omegah})
\begin{equation}
\frac{\nu_\sigma (\omega)}{\nu_0} = 1
- \left\{
\begin{array}{ll}
\frac{W_2}{\Omega_H}\frac{\pi^2}{\cosh^2
\left(\frac{\omega+ \sigma E^* }{\Omega_H}\right)},
& \left|\omega+\sigma E^*\right| \lesssim \Omega_H;\\
4 \pi^2 \frac{ W_2}{ \Omega_H}
\exp\left(- 2\frac{\omega+ \sigma E^* }{\Omega_H}\right)
+ \left(\frac{W_2}{\omega+ \sigma E^* }\right)^2
\ln\left(\frac{\Omega}{\omega+ \sigma E^* }\right)^2
,
& \left|\omega+\sigma E^*\right| \gg \Omega_H.
\end{array}\right.
\label{as2}
\end{equation}

\section{Experiments on \mbox{$Al$} films}
\label{sec:exp}

The theoretical study presented in this paper was inspired by the
experimental work of Wu, Williams, and Adams\cite{WWA}.  These authors
studied tunneling anomalies in ultra-thin (about $4nm$ thick) $Al$
films, which were driven into paramagnetic state by parallel magnetic
field $H > H_{\parallel} \simeq 4.8 T$.  Both zero bias anomaly and
anomalies at biases close to Zeeman splitting $E_Z$ were observed.
The authors attempted to fit the experimental results by the theory of
Ref.~\onlinecite{AA}, developed for normal metals and superconductors
at $T>T_c$.  The agreement appeared to be reasonable with one
important exception: the positions of the satellite singularities
$E^{**}$ was lower in energies than that predicted by
Ref.~\onlinecite{AA}: experimentally it was fitted as 
\begin{equation}
E^{**} \simeq
E_Z -E_0;\quad  E_0=0.17meV.
\label{experiment}
\end{equation}  

In our theory $eV_s = E^* < E_Z$, see
Eq.~(\ref{E^*}), and the discrepancy is reduced even though it does
not disappear. For instance, minimal value of $E^*$ corresponds to the
phase transition point $E_Z=\sqrt{2}\Delta$ and, according to
Eq.~(\ref{E^*}), equals to $\Delta (\sqrt{2}+1)/2 \simeq 0.47 meV$,
since $\Delta \simeq 0.39 meV$.  Experimental value $eV_s \simeq 0.38
meV$ is about $20\%$ (rather than $33\%$ in comparison with
Ref.~\onlinecite{AA}) smaller.  We do not have enough data to
speculate about possible sources of this discrepancy. More experiments
with serious quantitative analysis are needed to verify the present
theory.  Nevertheless, it may be worthwhile to briefly discuss here
how the other experimental findings of Ref.~\onlinecite{WWA} compare
with the theoretical conclusions.

In Ref.~\onlinecite{WWA}, authors presented and discussed tunneling
conductance $G(V,H)$ as a function of the bias voltage and magnetic
field for two samples. Both samples were granular $Al$ films about
$4nm$ thick.  Sheet resistance were different $R_{\Box}^{(1)} \simeq
4.2 K \Omega$, $R_{\Box}^{(2)} \simeq 2.0 K \Omega$.  For both samples
dips of the tunneling conductance at $V= \pm V_s$ were observed. The
widths at half minimum(WHM) of these dips for both samples were about
$0.15 \div 0.2 meV$, while the depths were found to be quite
different: $|\delta G/G|^{(1)} \simeq 0.12$ and $|\delta G/G|^{(2)}
\simeq 0.05$.

One can interpret these experimental results in two different
ways. The first interpretation is based on the assumption that the
granular structure of the films is irrelevant, and they can be
approximated as homogeneous 2D objects. Given the superconducting gap
$\Delta \simeq 0.39 meV$, Zeeman splitting $E_Z \simeq 0.57 meV$ at
magnetic field $H=5T$, and $R_{\Box}$, one can use Eq. (\ref{Wd}) to
determine $W_2$: $W_2^{(1)} \simeq 0.03 meV$; $W_2^{(2)} \simeq
0.015 meV $.  Since WHM of the anomaly should be compared with
approximately $2W_2$, see Fig.~\ref{Fig:5.1}c, we have not great but
reasonable agreement, especially for the first sample.  However the
law $W_2 \propto \ln(g)/g$ from Eqs.~(\ref{Wd}) and (\ref{2d-dos})
seems to contradict the experiment.

The alternative interpretation is based on the approximation of weakly
connected $Al$ grains: in the first approximation we neglect the
coupling between the grains.  This allows to use $0D$ expression for
the width of the singularity $W_0$, see Eq.~(\ref{W_0}).  Given the
electron concentration in $Al$ $n=1.8 \times 10^{23} cm^{-3}$ and
their Fermi energy $E_F=11.8 eV$ \cite{MA}, bare DoS estimates as $\nu
\simeq 2 \times 10^{22} (eV cm^3)^{-1}$.  Assuming that the grains in
lateral directions have typical size $b \simeq 30nm$\cite{WWA}, and
that the film thickness is $a \simeq 4nm$, we can estimate the mean
level spacing $\bar{\delta} \simeq 1/(ab^2 \nu) \simeq 0.03
meV$. Substitution of this value of $\bar{\delta}$ into
Eq.~(\ref{W_0}) gives $W_0 \simeq 0.11 meV$. This is in a good
agreement with the experiment, since WHM at zero dimensions according
to Fig.~\ref{Fig:5.1}a should be compared with $2W_0 \simeq 0.22$.

In order to understand the substantial difference in amplitudes of the
tunneling anomalies for the two samples, let us discuss the effect of
coupling between the grains. This coupling results in a finite dwell
time $\tau_{dwell}$ which an electron spends in a given grain before
tunneling into a neighboring one.  We can determine $\tau_{dwell}$
from $D$-the constant of the diffusion at times bigger than
$\tau_{dwell}$ using the relation $D\simeq b^2/(2\tau_{dwell})$.
Given the sheet resistance $R_{\Box}^{(1,2)}$, DoS $\nu$, and the film
thickness $a$, one can estimate $D$ as $D^{(1)} \simeq 0.2 cm^2/sec$,
$D^{(2)} \simeq 0.4 cm^2/sec$. As a result
\[
\frac{\hbar}{\tau_{dwell}^{(1)}} \simeq 0.05 meV \quad
\frac{\hbar}{\tau_{dwell}^{(2)}} \simeq 0.1 meV .
\]

Now we can explain the difference in the depths of the anomalies in the
two samples assuming that $\hbar/\tau_d$ contributes to $\Gamma$ from
Eq.~(\ref{etatot}). One can see from Fig.~\ref{Fig:5.1}a that the dip
at $\Gamma =W_0/2$ is approximately twice as deep as the one at
$\Gamma=W_0$.  At the same time, WHMs in these two cases are close to
each other.

Note that Eq.~(\ref{mfd}) with $\Gamma=\hbar/\tau_{dwell}$ can be
justified only for $\tau_{dwell} W_d > \hbar$.  Theoretical
investigation of the crossover between 0D and 2D behavior in granular
films goes beyond the framework of this paper, though such a study can
be important for a quantitative discussion of experiments.

From the perpendicular critical field $H_{c \perp} \simeq 1.5T$ and
the dimensions of a grain one can estimate $1/\tau_H$ and find that it
is irrelevant for the experiment\cite{WWA}.  The same is correct for
spin-orbit scattering.  Recent studies of tunneling through $Al$
grains\cite{Ralph} show that the difference of the $g$-factor from 2
is very small not only in average, but also for a given orbital as
well. Both $\hbar/\tau_H$ and $\hbar/\tau_{so}$ are probably smaller
than $0.01 meV$ and much smaller than $\hbar/\tau_{dwell}$.

{\narrowtext
\begin{figure}[ht]
\vspace{0.4 cm}
\epsfxsize=7.7cm
\centerline{\epsfbox{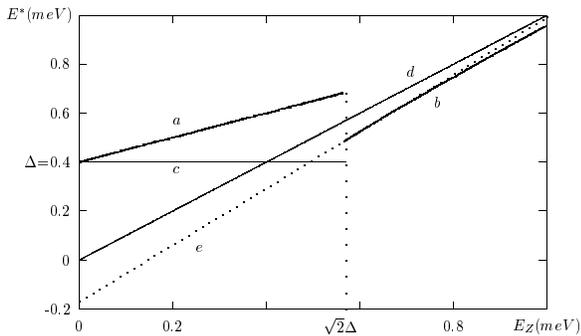}}
\vspace{0.35 cm}
\caption{
Position of the minimum in the DoS as the function of the Zeeman
splitting $E_Z$ for $\Delta\simeq 0.4 meV$: (a) Theoretical prediction
for the superconducting state\protect\cite{Fulde}, see
Fig.~\protect\ref{Fig:1}a;
(b) Our theoretical prediction, Eq.~(\protect\ref{E^*}) for the paramagnetic
state;
(c) Half distance between maxima in the DoS in the superconducting
state:
(d) $E^*=E_Z$ law predicted for the normal metal\protect\cite{AA};
(e) Approximation of Eq.~(\protect\ref{E^*}) by a straight line.
}
\vspace{0.15cm}
\label{Fig:6.1}
\end{figure}
}
\widetext

Let us now return to the discussion of the dip location.  Note that a
dip of DoS at finite bias exist in both superconducting and
paramagnetic states.  According to idealized Fig.~\ref{Fig:1}a in
superconducting side of the Clogston-Chandrasekhar phase transition
this anomaly is located at $eV=\Delta+E_Z/2$ (line ``a'' in
Fig.~\ref{Fig:6.1}).  However, probably due to the smearing of the DoS
singularities, experimentally the minimum was found in the middle
between two peaks in the DoS, {\em i.e.} at $eV\simeq\Delta$ (line
``c'' at Fig.~\ref{Fig:6.1}).  As it was already mentioned, the
experimentally found position of the singularity is lower than our
theoretical prediction (\ref{E^*}).  In fact there were no jump in
$E^*$ observed at the point of the first order phase transition.  This
discrepancy may be due to the inhomogeneous broadening of the
transition-different granulars may have slightly different $\Delta$.
Another possibility is illustrated on Fig.~\ref{Fig:6.1}.  In the
interval of magnetic fields where the measurements were done the
theoretical dependence Eq.~(\ref{E^*}) (line ``b'' at
Fig.~\ref{Fig:6.1}) can be approximated by 
\begin{equation}
E^* \approx r E_Z-0.17 meV
\label{theoryapprox}
\end{equation} 
with numerical factor $r \simeq 1.15 $ is slightly larger than $1$
(line ``e'' at Fig.~\ref{Fig:6.1}).  Comparing
Eq.~(\ref{theoryapprox}) with the experimental fit (\ref{experiment}),
we see that the theory would agree with experiment very well if we
assume that actual $g$-factor is smaller than its bare value, {\em
i.e.} $g_L=2/r\approx 1.72$.
 
\section{Conclusion}

This paper is devoted to the anomalies of the tunneling density of states of
low dimensional ($d=0,1,2$) superconductors in external magnetic field. We
concentrated on the  Clogston - Chandrasekhar (CC) phase
transition, {\em i.e.} the destruction of the superconductivity by the
magnetic field by virtue of the Zeeman splitting. As
a result  normal paramagnetic state of electrons is formed.

The main conclusion we can draw from our study of CC state is, that
despite this state being normal (mean-field superconducting order parameter
vanishes), it is drastically different from a usual normal
metal with some attractive interaction. The latter state appears,
{\em e.g.}, in a superconductor at temperatures higher than the
transition temperature $T_c$.  The difference becomes apparent when
one studies excited states rather than those closed to the ground
state.

Superconducting fluctuations in a usual normal disordered metal were
known to contribute to the zero-bias tunneling anomaly as well as to
Zeeman anomalies at the bias $eV$ equal to Zeeman splitting\cite{AA}.
However, these contributions (effects of the interaction in Cooper
channel) are similar or weaker than effects of Coulomb repulsion of
electrons, unless the system is anomalously close to the transition,
{\em i.e.}  it is not in Levanyuk-Ginzburg region.  This means that
the effects of superconducting fluctuations can be taken into account
perturbatively almost everywhere, (except the very vicinity of the
transition temperature) if the system is not too dirty. The
perturbative approach (expansion in inverse powers of the conductance
$g$) is valid as long as all of the characteristic length scales
involved into the problem do not exceed the localization length
$L_{loc}$.

Tunneling anomalies in CC normal state studied by us are quite
different. First of all, its position $eV = E^*$, see Eq.~(\ref{E^*}),
is different from the Zeeman splitting $E_Z$. However, what is more
important, the perturbative corrections to the density of states $\nu
(\omega )$ are much singular at $\omega $ close to $ E^*$ than the
same order in $g^{-1}$ corrections in usual normal metals. Because of
this, the perturbative approach fails at parametrically wider energy
interval $| eV - E^*| \leq W_d$ around the singular bias than that for
the normal metal.  Using Eqs.~(\ref{Wd}), one can check that the
length scale $L_W$ which corresponds to $W_d$ is much less than
$L_{loc}^{(d)}$, provided $L_{loc}$ exceeds superconducting coherence
length $\xi$.  Indeed, since the localization length can be estimated
as $L_{loc}^{(1)} \simeq D\nu^{(1)}$ and $L_{loc}^{(2)} \simeq
l\exp\left(D\nu^{(2)}\right) $, (where $l$ is the mean free path, $D$
is the diffusion coefficient, and $\nu^{(d)}$ is $d$ - dimensional
DoS) the characteristic spatial scale corresponding to the
singularity, $L_{W_d}$, can be written as
\begin{eqnarray*}
&\frac{L_{W_1}}{L_{loc}^{(1)}} \simeq \left(
\frac{\xi}{L_{loc}^{(1)}}\right)^{2/3}&\\
&\frac{L_{W_2}}{L_{loc}^{(2)}} \simeq \frac{\xi}{L_{loc}^{(2)}}
\sqrt{\ln\left(\frac{L_{loc}}{l}\right)}&.
\end{eqnarray*}

The fact that $W_d \gg D/L_{loc}^2$ makes it necessary and also
possible to go beyond the perturbation theory -- one has to sum only
most diverging terms, and it is allowed to neglect usual weak
localization and interaction corrections. It turns out to be possible
to sum directly whole series of the perturbation theory and thus
determine the shapes of the singularities at all dimensions.

The singularities are characterized by their widths $W_d$ given by
Eqs.~(\ref{W_0}) and (\ref{Wd}). For zero dimensional grains our
theory predicts a hard gap in the density of states with a given spin
direction, centered at $\omega = E^*$.  For one dimensional wire the
shape becomes universal (independent on $\nu ^{(1)}$ and $D$) when
energy is measured in units of $W_1$, see Eq.~(\ref{1d-dos}). It means
that the depth of the anomaly is universal. In the case of two
dimensional film the depths of the anomaly is not universal and
behaves as the inverse logarithm of the conductance, see
Eq.~(\ref{2d-dos}).

The reason for the effects of superconducting fluctuations in CC metal
to be dramatically enhanced in comparison with the usual case is the
presence the pole-like singularity in the correlation function of
these fluctuations.  This pole at a finite frequency appears due to
the fact that CC transition is of the first order. Contrarily, the
temperature driven transition from superconductor to normal metal is
of the second order, and in a usual normal state the correlator of the
superconducting fluctuations is a smooth function of the frequency,
{\em i.e.} any superconducting excitation decay very rapidly. We
believe that the strong anomalies of the excitation spectrum at finite
energies is a generic feature of any state created as a result of a
first order quantum phase transition.

\acknowledgements
Discussions with A.I. Larkin and B.Z. Spivak are acknowledged with
gratitude.


\begin{references}
\bibitem{footnote} Numerical coefficient in the formula for $W_1$ is
different from that in Ref.~\onlinecite{Igor} because of the algebraic
error in this reference.
\end{references}
\end{document}